\documentclass[12pt]{article}
\pdfoutput=1
\usepackage{jhep-mod} 
\usepackage{bm}
\usepackage{amssymb}
\usepackage{pifont}
\newcommand{\cmark}{\ding{51}}%
\newcommand{\xmark}{\ding{55}}%
\title{Semiclassical energy conditions for quantum vacuum states}
\author{Prado Mart\'in--Moruno \textmd{and} Matt Visser}
\affiliation{School of Mathematics, Statistics, and Operations Research, \\
Victoria University of Wellington, PO Box 600, Wellington 6140, New Zealand}
\emailAdd{prado@msor.vuw.ac.nz}
\emailAdd{matt.visser@msor.vuw.ac.nz}
\abstract{
We present and develop several nonlinear energy conditions suitable for use in the semiclassical regime. In particular, we consider the recently formulated ``flux energy condition" (FEC), and the novel ``trace-of-square" (TOSEC) and ``determinant" (DETEC) energy conditions.
As we shall show, these nonlinear energy conditions behave much better than the classical linear energy conditions in the presence of semiclassical quantum effects. Moreover, whereas the quantum extensions of these nonlinear energy conditions seem to be quite widely satisfied as one enters the quantum realm, at least for quantum vacuum states, analogous quantum extensions  are generally not useful for the linear classical energy conditions.

\bigskip
\noindent
7 June 2013; 24 July 2013; 7 August 2013; \LaTeX-ed \today
}
\begin{document}
\maketitle
\newcommand {\apgt} {\ {\raise-.5ex\hbox{$\buildrel>\over\sim$}}\ }
\newcommand {\aplt} {\ {\raise-.5ex\hbox{$\buildrel<\over\sim$}}\ }
\def\d{{\mathrm{d}}}
\def\x{\bm{x}}
\def\p{\bm{p}}
\def\0{\bm{0}}
\def\n{\bm{n}}
\def\vv{\bm{v}}
\def\u{\bm{u}}
\def\w{\bm{w}}
\def\bpi{\bm{\pi}}
\def\blambda{\bm{\lambda}}
\def\O{{\mathcal{O}}}
\def\P{{\mathcal{P}}}
\def\C{{\mathcal{C}}}
\def\E{{\mathcal{E}}}
\def\R{{\mathbb{R}}}
\def\bb{{b_H'' }}
\def\bbb{{b_H''' }}
\def\x{\bm{x}}
\def\p{\bm{p}}
\def\0{\bm{0}}
\def\n{\bm{n}}
\def\v{\bm{v}}
\def\u{\bm{u}}
\def\w{\bm{w}}
\def\bpi{\bm{\pi}}
\clearpage
\section{Introduction}
\label{S:intro}

The classical energy conditions have in the past  been proven to be of undoubted practical usefulness for extracting general characteristics of space-time without restricting attention to a particular matter model.  Results along these lines include the well known singularity theorems, the laws of black hole thermodynamics, (specifically the area increase theorem), and the positivity proof for the ADM mass~\cite{Matt}. Nevertheless, those classical energy conditions are not fundamental physics, and will not be fulfilled in completely general situations~\cite{twilight}. The strong energy condition already fails miserably in the classical regime when describing our Universe, both during inflation and during the current accelerated phase~\cite{galaxies, epoch, epoch-mg8}. Moreover, violations of \emph{all} the classical energy conditions, even \emph{unbounded} violations, generically appear when considering non-vacuum quantum states in semiclassical physics~\cite{twilight, early, Visser:1994, gvp1, gvp2, gvp3, 
gvp4, gvp5, Flanagan:1996, q-anec, volume-aec}. (Non-vacuum squeezed states are particularly prone to such effects.) Therefore, those classical linear energy conditions should clearly be replaced by other assumptions whenever quantum effects cannot be ignored.

The nonlocal time-integrated Ford--Roman quantum inequalities are one way of side-stepping this issue, and developing some significant constraints on the semiclassical stress-energy tensor~\cite{Ford:1990, Ford+Roman, Ford:1996, Ford+Roman+collaborators, others, Flanagan:1997, Fewster:1998, Ford:1999, Fewster:1999a, Fewster:1999b, Teo:2002, Fewster:2007, Abreu:2008, Abreu:2010}. Those inequalities have led to the so-called ``quantum interest conjecture", which attempts to more precisely quantify the size of the quantum-induced violations of the classical energy conditions. 

On the other hand,  a new classical point-wise energy condition has recently been presented~\cite{Abreu:2011}, and has already been demonstrated to be useful for obtaining classical entropy bounds. This is the so-called ``flux energy condition" (FEC)~\cite{Abreu:2011}. Moreover, as pointed out in~\cite{Prado:2013}, this FEC is rather better behaved than the usual energy conditions (NEC, WEC, SEC, DEC),  and is genuinely different. In reference~\cite{Prado:2013} a quantum version of the FEC (QFEC) has been presented. This QFEC is satisfied for naturally occurring quantum vacuum states under rather general but non-tautological conditions, and provides useful semiclassical constraints on the stress-energy. 

Following in this spirit, we shall consider further nonlinear (quadratic and quartic) constraints on the stress-energy and their possible quantum generalization. For this purpose we will study in deeper detail the characteristics and applications of the FEC and QFEC, and moreover, we will present two novel energy conditions which entail the consideration of quantities such as the determinant of the stress-energy, or the trace of the square of the stress energy. Despite not having quite so clear a physical interpretation as the FEC, we shall show that these conditions and their quantum generalizations could  (at least in principle) be very useful in a broader range of situations than the classical energy conditions.

\clearpage
The present paper is structured as follows: In section~\ref{sFEC} we shall summarise the definitions and characteristics of the FEC and QFEC. We will introduce the commonly used type I--IV classification of Hawking and Ellis~\cite{Hawking-Ellis} to consider \emph{necessary} and \emph{sufficient} conditions for the FEC/QFEC to be satisfied, in subsections~\ref{ssT} and \ref{ssSS}, respectively. In sections~\ref{sDETEC} and \ref{sTOSEC} we shall present the ``determinant energy condition" (DETEC) and the ``trace-of-square energy condition" (TOSEC), respectively, as well as their quantum generalisations. The fact that such quantum generalisations are definitely nontrivial can be noted by studying the fulfilment/violation of a similar quantum generalisation of the usual linear energy conditions, which will be introduced in section~\ref{sQECs}.

After formulating these various candidate nonlinear energy conditions we shall test them against several specific situations. We shall investigate the renormalised stress-energy tensor in the Casimir vacuum in section~\ref{sCasimir}, for the massless minimally coupled scalar in a 3+1 Schwarzschild geometry in section~\ref{sSchw}, and for generic 1+1 QFTs in section~\ref{s11} --- paying particular attention to the differences between the Boulware, Hartle--Hawking, and Unruh quantum vacuum states. Compared to the usual energy conditions, all of these nonlinear energy conditions are much better behaved as one enters the quantum realm --- certainly when evaluated on standard quantum vacuum states.
All these issues will be discussed in section~\ref{sD}.
It must be noted that in this paper we shall work in an orthonormal basis even in curved spacetime except where explicitly noted.

\section{Flux energy condition}\label{sFEC}

The energy-momentum 4-flux seen by a timelike observer of 4-velocity $V^a$ is
\begin{equation}
F^a = - \langle T^{ab}\rangle \, V_b. 
\end{equation}
We will now use this flux 4-vector to summarise some results about the FEC and QFEC.
(To unify the notation, in the classical realm we shall take $\langle T^{ab}\rangle$ to just denote the classical stress energy $T^{ab}$, while in the semiclassical realm we shall take this to denote a renormalized expectation value; in all the specific cases we consider it in fact denotes a renormalized vacuum expectation value.)

\subsection{FEC}\label{ssFEC}

The FEC is simply the demand that the flux 4-vector of any timelike observer be timelike or at worst null~\cite{Abreu:2011}:
\begin{equation}\label{FEC}
F^a F_a \leq 0.
\end{equation}
Because there is no demand that the flux 4-vector be future pointing this is a strictly weaker condition than the DEC, which requires the WEC, $\langle T^{ab}\rangle  \, V_a V_b \geq 0$, in addition to the FEC. That is, the FEC dose not restrict the sign of the energy density.
A useful way to re-write the FEC is using the tensor $[-T^2]^{ab}= - \langle T^{ac}\rangle\,g_{cd} \,\langle T^{db}\rangle$. Thus, Equation~(\ref{FEC}) is equivalent to
\begin{equation}\label{FEC2}
 [-T^2]^{ab} \; V_a V_b \geq 0,
\end{equation}
and the FEC can be understood as a variant of the WEC applied to sign reversed square of the stress-energy tensor~\cite{Prado:2013}.

The implications of the FEC can be studied adopting an orthonormal basis and performing rotations in the spatial part until  $T^{ij}$ is diagonal. That is, without loss of generality, we can choose
\begin{equation}
\label{E:diag}
\langle T^{ab} \rangle =
 \left[\begin{array}{c|ccc}\rho&f_1&f_2&f_3\\ \hline f_1&p_1&0&0\\f_2&0&p_2&0\\f_3&0&0&p_3\end{array}\right]\!;
\quad
V^a = \gamma\left(1;\beta_1,\beta_2,\beta_3\right); 
\quad
F^a = \gamma\big(\rho-\vec \beta\cdot\vec f ; \;  f_i-p_i\beta_i\big),
\end{equation}
where no summation on repeated indices is implied, and $\sum\beta_i^2 < 1$.
We can then write
\begin{equation}
[-T^2] ^{ab} =
 \left[\begin{array}{c|ccc}\rho^2-\vec f\cdot\vec f&(\rho-p_1)f_1\;&\;(\rho-p_2)f_2\;&\;(\rho-p_3) f_3\\ 
 \hline 
 (\rho-p_1)f_1&-p_1^2+f_1^2&f_1 f_2&f_1 f_3\\
 (\rho-p_2)f_2&f_2 f_1&-p_2^2+f_2^2&f_2 f_3 \\
 (\rho-p_3)f_3&f_3 f_1&f_3 f_2&-p_3^2+f_3^2\end{array}\right],
\quad
\end{equation}
which, introduced into equation~(\ref{FEC2}), or alternatively proceeding directly from the ultimate equation in (\ref{E:diag}), leads to the conditions
\begin{equation}\label{FECbeta}
\gamma^2\Big( \big[\rho-\vec\beta\cdot\vec f\;\big]^2  - {\textstyle\sum}\left[ f_i-p_i\beta_i\right]^2\Big) \geq 0,
\end{equation}
for any combination of $\beta_i$ such that $\sum\beta_i^2 < 1$.

By taking the limits $\beta_i \rightarrow\pm1$ and $\beta_i\rightarrow0$ in~(\ref{FECbeta}), one can obtain the following seven constraints:
\begin{equation}
\label{E:seven}
(\rho \pm f_i)^2 - (p_i\pm f_i)^2 - \sum_{j\neq i} f_j^2 \geq 0; 
\qquad
\rho^2 - \vec f \cdot \vec f \geq 0,
\end{equation}
which must be \emph{necessarily} fulfilled for the FEC to be satisfied.
(Sometimes the seventh constraint is redundant, for instance, if any one of the $f_i$ is zero.) As we shall show in~\ref{ssSS}, these constraints are also sufficient in most cases of specific interest.

\subsection{QFEC}\label{ssQFEC}
The QFEC is a condition requiring the flux 4-vector not to be ``excessively spacelike''. Specifically, as formulated in~\cite{Prado:2013}, the weak QFEC states that the the flux should either be non-spacelike or have a norm bounded from above. Moreover, a strong version of this condition has also been introduced taking into account that a natural bound on the stress-energy components should depend both on $\hbar$ (in order to recover the FEC for classical systems) and on general characteristics of the system under consideration (such as its size and state of motion). Thus, the strong QFEC is taken to be~\cite{Prado:2013}
\begin{equation}\label{QFEC}
F^aF_a \leq \zeta \; (\hbar N/L^4)^2\; (U_a\,V^a)^2,
\end{equation}
where $\zeta$ is a generic positive number of order unity, (not necessarily always the \emph{same} positive number), $N$ is the number of quantum fields under consideration, $L$ is a characteristic distance associated with the system, $U^a$ is the system 4-velocity, and $V^a$ is the observer 4-velocity.
(Thus, by definition, the QFEC makes sense only in those situations for which there is a well-defined notion of system size and 4-velocity.)

Taking into account~(\ref{E:diag}), the QFEC can be written as
\begin{equation}\label{QFECbeta}
 \big[\rho-\vec\beta\cdot\vec f\;\big]^2  - {\textstyle\sum}\left[ f_i-p_i\beta_i\right]^2 \geq -\zeta\;(\hbar N/L^4)^2,
\end{equation}
with $\sum\beta_i^2 < 1$.
Thus, following a similar procedure as that applied in the previous section, we find \emph{necessary} conditions for the strong QFEC to hold. These are:
\begin{equation}
\label{E:qfec2}
(\rho \pm f_i)^2 - (p_i\pm f_i)^2 -\sum_{j\neq i} f_j^2 \geq - \zeta\;(\hbar N/L^4)^2;
\end{equation}
\begin{equation}
\label{E:qfec3}
\rho^2 - \vec f \cdot \vec f \geq - \zeta\;(\hbar N/L^4)^2.
\end{equation}
In some situations the quantity $\zeta$ can be set to unity if desired. (See~\ref{ssSS} for more details on this point.)

\subsection{Stress-energy types}\label{ssT}
In order to consider the fulfillment (or violation) of the FEC in some particular cases, let us consider the Hawking--Ellis classification of stress-energy tensors~\cite{Hawking-Ellis}.  
Hawking and Ellis classify the stress-energy into one of four possible types depending on the extent to which they can be diagonalised by some arbitrary \emph{Lorentz} transformation:
\begin{equation}
\langle T^{ab} \rangle =   (\hbox{canonical type})^{cd} \; L_c{}^a\; L_d{}^b.
\end{equation}
The four canonical types can be taken to be:
\begin{equation}\label{E:HE12}
 \left[ \begin{array}{c|ccc} \rho&0&0&0\\ \hline0&p_1&0&0\\0&0&p_2&0\\0&0&0&p_3 \end{array} \right];
 \qquad\qquad\qquad
  \left[ \begin{array}{cc|cc} \mu+f& f&0&0\\  f&-\mu + f &0&0\\ \hline0&0&p_2&0\\0&0&0&p_3 \end{array} \right];
 \end{equation}
 and
 \begin{equation}\label{E:HE34}
  \left[ \begin{array}{ccc|c} 
 \rho\;&\; {1\over\sqrt{2}}f\;&\;{1\over\sqrt{2}}f&0\\ 
 {1\over\sqrt{2}}f\;&\;-\rho+ f\;&\;0&0\\
 {1\over\sqrt{2}}f\;&\;0\;&\;-\rho-f &0\\ \hline
 0&0&0&p_3 \end{array} \right];
 \qquad
 \left[ \begin{array}{cc|cc} \rho&f&0&0\\ f&-\rho&0&0\\ \hline 0&0&p_2&0\\ 0&0&0&p_3 \end{array} \right]. 
\end{equation}
We have chosen a slightly different presentation of the type II, III, and IV stress tensors which we find more useful for our purposes. Note that as presented here, these are all sub-cases of our equation (\ref{E:diag}). 

This classification into types I--IV is closely related to the study of eigenvalues and eigenvectors of the mixed tensor $\langle T^a{}_b\rangle = \langle T^{ac}\rangle\, g_{cb}$.  It is the fact that the mixed tensor $\langle T^a{}_b\rangle$ is \emph{not symmetric} that is ultimately the source of the non-trivial types II--IV. The pattern of the Lorentz invariant eigenvalues of $\langle T^a{}_b\rangle$ for these 4 canonical types is:
\begin{equation}
[-\rho\,|\,p_1,p_2,p_3]; \qquad [-\mu,-\mu\,|\,p_2,p_3]  \qquad [-\rho,-\rho,-\rho\,|\,p_3]; \qquad [ -\rho\pm i f    \,|\,p_2,p_3].
\end{equation}
(The eigenvalues of $\langle T^{ab}\rangle$, with both indices up, are neither Lorentz invariant, nor interesting, nor pleasant to deal with.)

While type III and type IV stress tensors do not seem to arise from simple classical field theories~\cite{Hawking-Ellis}, as we shall soon see, the situation with respect to renormalized semiclassical stress-energy tensors is much more subtle. 
For these four canonical types the reversed square of the stress tensor, ($[-T^2]^{ab}$), takes on the forms: 
\begin{equation}
 \left[ \begin{array}{c|ccc} \rho^2&0&0&0\\ \hline0&-p_1^2&0&0\\0&0&-p_2^2&0\\0&0&0&-p_3^2 \end{array} \right];
 \qquad
  \left[ \begin{array}{cc|cc} \mu^2+2\mu f& 2\mu f&0&0\\  2\mu f&-\mu^2 + 2\mu f &0&0\\ \hline0&0&-p_2^2&0\\0&0&0&-p_3^2 \end{array} \right];
\end{equation}
\begin{equation}
  \left[ \begin{array}{ccc|c} 
 \rho^2-f^2\;&\; {1\over\sqrt{2}}f(2\rho - f)\;&\;{1\over\sqrt{2}}f(2\rho+f)&0\\ 
 {1\over\sqrt{2}}f(2\rho  - f)&\;-\rho^2-{1\over2} f^2+2 \rho f\;&\;{1\over2}f^2&0\\
 {1\over\sqrt{2}}f(2\rho + f)\;&\;{1\over2}f^2\;&\;-\rho^2-{1\over2}f^2-2 \rho f&0\\ \hline
 0&0&0&-p_3^2 \end{array} \right];  
 \end{equation}
 and
 \begin{equation}
 \left[ \begin{array}{cc|cc} \rho^2-f^2& 2\rho f&0&0\\ 2\rho f&-(\rho^2-f^2)&0&0\\ \hline 0&0&-p_2^2&0\\ 0&0&0&-p_3^2 \end{array} \right]. 
\end{equation}
For types I, II, and IV the reversed square is again of the same \emph{form} as the original tensor.
For type III the reversed square is qualitatively different from the original tensor.

In terms of these four canonical types the \emph{necessary constraints} for the FEC to be satisfied reduce to:
\begin{itemize}
\item 
type I: \qquad
$ p_i^2 \leq \rho^2$.\\
This may or may not be satisfied. As we will see in the next subsection, these are also sufficient conditions for the FEC to be satisfied; therefore, the FEC provides nontrivial constraints in this case.
\item 
type II: \qquad
$\mu f \geq 0$; \qquad
$p_i^2 \leq \mu^2  +2\mu f$.\\
This may or may not be satisfied; thus, in principle, the FEC would in this situation provide a nontrivial constraint. Nevertheless, a more careful analysis is needed that considers the sufficient conditions.
\item
type III:  In this case we have
\begin{equation}
-{3\over2} f^2 + 2\rho f \pm\sqrt{2}(2\rho f - f^2) \geq 0;  
\quad 
-{3\over2} f^2 - 2\rho f \pm\sqrt{2}(2\rho f + f^2) \geq 0;   
\end{equation}
and
\begin{equation}
\quad \rho^2-f^2-p_3^2\geq 0. 
\end{equation}
The only way all these constraints can be satisfied is if $f=0$, 
in which case it degenerates to a special case of type I.
Thus, a ``genuine'' type III tensor with $f\neq0$ can never satisfy the FEC.\\
\item
type IV: \qquad 
$\rho f\leq 0$; \qquad $\rho f \geq 0$; \qquad $p_i^2 \leq \rho^2-f^2$; \qquad $f^2\leq \rho^2$.\\
These constraints lead to $\rho f =0$. If $f=0$ this implies degeneration to a special case of type I,
while if $\rho=0$  the type IV tensor degenerates to zero.
Therefore, a ``genuine'' type IV tensor with $\rho\neq0\neq f$ can never satisfy the FEC.
\end{itemize}
Overall, considering only the necessary constraints~(\ref{E:seven}), we have been able to conclude that the FEC cannot be satisfied for stress tensors which are either genuine type III or genuine type IV.

\subsection{Sufficient conditions}\label{ssSS}

Up to know we have only considered necessary constraints for the FEC to be satisfied. 
Thus, in cases where this constraints are fulfilled, we cannot yet extract any definitive conclusion.
In this section, we will obtain sufficient conditions for the FEC and QFEC to be satisfied by a given stress energy tensor for any observer.

\subsubsection{1+1 dimensions}

The 1+1 dimensional case is a highly symmetric situation in which the necessary conditions~(\ref{E:seven}) are also sufficient. In order to understand this behaviour let us consider the FEC through equation~(\ref{FECbeta}) which, in this case, leads to
\begin{equation}\label{feq11}
(\rho-\beta f)^2-(f-\beta p)^2\geq0  \quad\implies\quad |\rho-\beta f|\geq|f-\beta p|,
\end{equation}
where $\beta^2<1$. As~(\ref{feq11}) is linear in $\beta$, one can argue that the necessary conditions for the FEC to be satisfied can be found by considering the limit values of $\beta$, that is $\beta\rightarrow\pm1$. Taking also into account the case $\beta=0$, to be sure that no additional information is provided when the parameter vanishes, those constraints are precisely those given in~(\ref{E:seven}).

Regarding the QFEC, it must be noted that the previous argument would not be generally valid if we introduce an explicit bound in the manner of equation~(\ref{QFECbeta}). That is because we are then not able to simplify the square as in~(\ref{feq11}) to obtain an expression which is linear in $\beta$. Therefore, it seems more natural to simply study the weak QFEC in generic situations. On the other hand, in the particular case in which $f=0$, equation~(\ref{QFEC}) reduces to
\begin{equation}
\rho^2-\beta^2p^2\geq-\zeta\; \left(\hbar N/L^2\right)^2,
\end{equation}
which is linear in the quantity $\beta^2$. Thus, we can be sure that it will be fulfilled for any allowed value of $\beta$ provided
\begin{equation}
\rho^2-p^2\geq-\zeta\;\left(\hbar N /L^2\right)^2.
\end{equation}

\subsubsection{Spherically symmetric spacetimes}

The situation that we have shown in 1+1 dimensions is specific to that highly symmetric space. In general, one cannot extract sufficient conditions which assure the fulfillment of the FEC in 1+3 dimensional spacetimes without restricting to particular cases, even when considering a spherically symmetric situation.
In order to see that, let us consider a stress energy tensor with the mentioned symmetry. This is:
\begin{equation}\label{Tss}
\langle T^{ab}\rangle    =\left[ \begin{array}{cc|cc} \rho&f&0&0\\  f&p_r&0&0\\ \hline 0&0&p_t&0\\0&0&0&p_t \end{array} \right].
   \end{equation}
The \emph{necessary} conditions in this case are:
\begin{equation}
\rho^2-p_r^2\pm2f(\rho-p_r)\geq0 \qquad\hbox{and}\qquad \rho^2-p_t^2-f^2\geq0.
\end{equation}
Whereas~(\ref{FECbeta}) can be re-written as
\begin{equation}\label{E:sss}
\hbox{FEC:}\qquad  (\rho-\beta_1 f)^2 - (f-\beta_1 p_r)^2 -\left( \beta_2^2+\beta_3^2\right) p_t^2 \geq 0.
\end{equation}
Using $\sum \beta_i^2 \leq 1$ we have
\begin{equation}\label{E:226}
\hbox{LHS} \geq \beta_1^2\left[\rho^2+f^2-p_r^2\right]  -2\beta_1 f(\rho-p_r) +\left(\beta_2^2+\beta_3^2\right)\left(\rho^2-p_t^2\right)-f^2.
\end{equation}
The only way to absolutely guarantee that this quantity is larger or equal than zero for any $\beta_i$ is by requiring that each addend, with a distinct power of $\beta_i$, be nonnegative. The resulting constraints would be, of course, \emph{sufficient} for the FEC to be satisfied. Nevertheless, they are so restrictive, that would they never be fulfilled. (Note, for example, that they would require $f^2<0$.)
Thus, it is not possible to extract general sufficient conditions which are not too restrictive, without assuming some relation between the different quantities, which would simplify~(\ref{E:226}). For this reason, we shall take into account that any stress energy tensor can be expressed in one of the four types given by equations~(\ref{E:HE12}) and (\ref{E:HE34}).

\subsubsection{Hawking and Ellis classification}

We now consider the stress energy tensor written in one of the four canonical forms. Moreover, it must be pointed out that in the spherical symmetric case, there are only three types of stress energy tensors compatible with that symmetry, types I, II, or IV. As we have pointed out, the specific type would depend on the spectrum of eigenvalues of the mixed tensor $\langle T^a{}_c\rangle =\langle T^{a b}\rangle g_{b c}$. In particular, taking into account the quantity
\begin{equation}
\Gamma=\left(\rho+p_r\right)^2-4f^2,
\end{equation}
we can conclude that the tensor is of  type I, II, or IV, if $\Gamma>0$, $\Gamma=0$, or $\Gamma<0$, respectively.

\paragraph{Type I.}
This is the case where $\langle T^a{}_c\rangle $ has one timelike eigenvector, with $\langle T^{ab}\rangle$ given by the matrix on the left of equation~(\ref{E:HE12}).
In this case equation~(\ref{FECbeta}) reduces to
\begin{equation}
{\rm FEC}: \qquad \rho^2-\beta_1^2p_1^2-\beta_2^2p_2^2-\beta_3^2p_3^2 \geq 0.
\end{equation}
Then, since $1\geq \sum \beta_i^2$, we have
\begin{equation}
\hbox{LHS} \geq \left(\beta_1^2+\beta_2^2+\beta_3^2\right)\rho^2-\beta_1^2p_1^2-\beta_2^2p_2^2-\beta_3^2p_3^2,
\end{equation}
whence
\begin{equation}
\hbox{LHS}\geq \beta_1^2\left(\rho^2-p_1^2\right)+\beta_2^2\left(\rho^2-p_2^2\right)+\beta_3^2\left(\rho^2-p_3^2\right).
\end{equation}
In view of this, conditions~(\ref{E:seven}),
\begin{equation}
\rho^2-p_i^2\geq0,
\end{equation}
are both \emph{necessary and sufficient} conditions (in this case) for the FEC to be satisfied. Moreover, for the spherically symmetric case, $p_2=p_3$, these three conditions reduce to two.

For a general study, independent of a particular basis, it is interesting to write these constraints using the eigenvalues.
Denoting by $\lambda_0$ the eigenvalue associated to the timelike vector, and by $\lambda_i$ the other eigenvalues, we can write
\begin{equation}
\lambda_0^2-\lambda_i^2\geq0.
\end{equation}

On the other hand, to consider a strong formulation of the QFEC, we assume, in the first place, that the fluxes measured by observers moving in the direction of the axes are bounded from below by $-(\hbar N/L^4)^2$. Thus,
\begin{equation}\label{qfectype1}
\rho^2-p_i^2\geq-\zeta\;(\hbar N/L^4)^2.
\end{equation}
Secondly, we note that this implies that the flux measured by any observer will be bounded from below by the same quantity. That is:
\begin{equation}
\hbox{QFEC:}\qquad
\rho^2-\beta_1^2p_1^2-\beta_2^2p_2^2-\beta_3^2p_3^2
\geq  -\left(\beta_1^2+\beta_2^2+\beta_3^2\right)\zeta(\hbar N/L^4)^2
\geq-\zeta(\hbar N/L^4)^2.
\end{equation}
Therefore, the constraints~(\ref{qfectype1}) are necessary and sufficient for the QFEC to be satisfied.

\paragraph{Type II.}
Let us now consider that $\langle T^{ab}\rangle$ is given by the matrix on the right of equation~(\ref{E:HE12}); thus,  $\langle T^a{}_c\rangle$ has a double null eigenvalue $\lambda=-\mu$.
In this case equation~(\ref{FECbeta}) can be written as
\begin{equation}\label{cfec1}
{\rm FEC}: \qquad (1-\beta_1^2)\mu^2+2(1-\beta_1)^2\mu f-p_2^2\beta_2^2-p_3^2\beta_3^2 \geq 0,
\end{equation}
which leads to
\begin{equation}
\hbox{LHS}\geq\beta_2^2(\mu^2-p_2^2)+ \beta_3^2(\mu^2-p_3^2)+2(1-\beta_1)^2\mu f.
\end{equation}
Thus, a set of \emph{sufficient conditions} for having the LHS $\geq0$ is
\begin{equation}\label{cfec3}
p_2^2\leq\mu^2,\qquad p_3^2\leq\mu^2, \qquad{\rm and}\qquad \mu f\geq0.
\end{equation}
These constraints are stronger than the necessary constraints~(\ref{E:seven}).

It will be useful for our study of the Unruh vacuum in Schwarzschild spacetime (section~\ref{sSchw}) to note that we could have chosen to express the type II stress energy tensor as
\begin{equation}\label{E:HE2b}
  \langle T^{ab}\rangle =\left[ \begin{array}{cc|cc} \mu-\bar f& \bar f&0&0\\  \bar f&-\mu - \bar f &0&0\\ \hline0&0&p_2&0\\0&0&0&p_3 \end{array} \right],
 \end{equation}
 with $\bar f=-f$.
In this case, we have
\begin{equation}\label{cfec1b}
{\rm FEC}: \qquad (1-\beta_1^2)\mu^2-2(1+\beta_1)^2\mu \bar f-p_2^2\beta_2^2-p_3^2\beta_3^2 \geq 0,
\end{equation}
which implies
\begin{equation}
\hbox{LHS}\geq\beta_2^2(\mu^2-p_2^2)+ \beta_3^2(\mu^2-p_3^2)-2(1+\beta_1)^2\mu \bar f,
\end{equation}
consistently leading to the following set of sufficient conditions \emph{sufficient conditions}:
\begin{equation}\label{cfec3b}
p_2^2\leq\mu^2,\qquad p_3^2\leq\mu^2, \qquad{\rm and}\qquad \mu \bar f\leq0.
\end{equation}

A formulation of the QFEC is in this case more subtle. Considering $T^{ab}$ given in the basis compatible with~(\ref{E:HE12}), the QFEC would be 
satisfied if
\begin{equation}
{\rm QFEC}:  \qquad (1-\beta_1^2)\mu^2+2(1-\beta_1)^2\mu f-p_2^2\beta_2^2-p_3^2\beta_3^2 \geq  -\zeta (\hbar N/L^4)^2.
\end{equation}
It can be noted that 
\begin{equation}
\hbox{LHS} \geq 2(1-\beta_1)^2\mu f+\beta_2^2\left(\mu^2-p_2^2\right)+\beta_3^2\left(\mu^2-p_3^2\right).
\end{equation}
Therefore, if we require
\begin{equation}\label{qfectype2}
\mu f\geq -{\zeta \over 4}(\hbar N/L^4)^2,\qquad \mu^2-p_i^2\geq-\zeta(\hbar N/L^4)^2,
\end{equation}
then the QFEC will be satisfied, since
\begin{equation}
\hbox{LHS} \geq -\left[\beta_2^2+\beta_3^2+(1-\beta_1)^2/2\right] \zeta(\hbar N/L^4)^2\geq-\bar\zeta(\hbar N/L^4)^2,
\end{equation}
and $\bar\zeta=2\zeta$ is a positive number of order unity, as is $\zeta$.
(In the following, we should not make any distinction between $\zeta$'s of order unity.)

\paragraph{Type III.} 
In this case the stress energy tensor can be written in an orthonormal basis as presented in the right side of equation~(\ref{E:HE34}), with $\langle T^a{}_b\rangle$ having triple eigenvalue associated with a null eigenvector. As we have already pointed out, a genuine type III
tensor can never satisfy the FEC.
Nevertheless, we can quantify the extent of violations of the FEC by considering equation~(\ref{QFECbeta}). This is:
 \begin{eqnarray}
 {\rm QFEC:}\quad&& \left(1-\beta_1^2-\beta_2^2\right)\rho^2-\left(1-\sqrt{2}(\beta_2-\beta_1)-2\beta_1\beta_2\right)f^2\nonumber\\
 &&-\sqrt{2}\left(2\beta_1+2\beta_2-\sqrt{2}\beta_1^2+\sqrt{2}\beta_2^2\right)\rho f-\beta_3^2p_3 \geq -\zeta(\hbar N/L^4)^2.
 \end{eqnarray}
 Then, the relevant quantity
 \begin{eqnarray}
 \hbox{LHS}&\geq&\beta_3^2\left(\rho^2-p_3^2\right)-\left(1-\sqrt{2}(\beta_2-\beta_1)-2\beta_1\beta_2\right)f^2\nonumber\\
 &&-\sqrt{2}\left(2\beta_1+2\beta_2-\sqrt{2}\beta_1^2+\sqrt{2}\beta_2^2\right)\rho f,
 \end{eqnarray}
 would be larger or equal the bound for any value of $\beta_i$, if the following \emph{sufficient} constraints are satisfied:
 \begin{equation}
\rho^2-p_3^2\geq-\zeta(\hbar N/L^4)^2,\qquad |f|\leq\zeta\hbar N/L^4,\qquad |\rho f|\leq\zeta(\hbar N/L^4)^2.
 \end{equation}

\paragraph{Type IV.}
In this case $\langle T^a{}_b\rangle$ has two complex eigenvalues, since $\Gamma<0$. As in the previous case, a genuine type IV tensor cannot satisfied the FEC.
However, the violations of this condition can again be quantified, now  taking into account equations~(\ref{QFECbeta}) and (\ref{E:HE34}). Thus, we have
\begin{equation}\label{feqIV}
{\rm QFEC}: \qquad \left(1-\beta_1^2\right)\rho^2-\left(1-\beta_1^2\right)f^2-4\beta_1\rho f-\beta_2^2p_2^2-\beta_3^2p_3^2 \geq -\zeta(\hbar N/L^4)^2,
\end{equation}
which implies
\begin{equation}
{\rm LHS}\geq
-\left(1-\beta_1^2\right)f^2-4\beta_1\rho f+\beta_2^2\left(\rho^2-p_2^2\right)+\beta_3^2\left(\rho^2-p_3^2\right).
\end{equation}
Then the QFEC  would be satisfied for any value of the parameters if
\begin{equation}
|f|\leq \zeta \hbar N/L^4,\qquad |\rho f|\leq  \zeta(\hbar N/L^4)^2,\qquad \rho^2-p_{2}^2\geq -\zeta (\hbar N/L^4)^2, 
\end{equation}
and
\begin{equation}
\rho^2-p_{3}^2\geq-\zeta(\hbar N/L^4)^2
\end{equation}
Recall that $\zeta$ denotes a generic number of order unity, and need not be the same number in the four inequalities above. 
Considering the two complex eigenvalues $\lambda$ and $\bar\lambda$, (and two real eigenvalues $\lambda_2$ and $\lambda_3$), the constraints for the QFEC are:
\begin{eqnarray}\label{qfecIV}
&&|{\rm Im}(\lambda)|\leq \zeta\hbar N/L^4;\qquad |{\rm Re}(\lambda){\rm Im}(\lambda)|\leq \zeta (\hbar N/L^4)^2; \nonumber\\
&&\left[{\rm Re}(\lambda)\right]^2-\lambda_{2}^2\geq-\zeta(\hbar N/L^4)^2; \qquad
\left[{\rm Re}(\lambda)\right]^2-\lambda_{3}^2\geq-\zeta(\hbar N/L^4)^2.
\end{eqnarray}

\section{Determinant energy condition}\label{sDETEC}

The DETEC will be taken to be
\begin{equation}
\det \langle T^{ab}\rangle \geq 0.
\end{equation}
In an orthonormal frame where the spatial stress has been diagonalized (\emph{cf} (\ref{E:diag})) this condition becomes
\begin{equation}
\det\langle T^{ab}\rangle = \rho\, p_1 p_2 p_3 - \left[ f_1^2 \, p_2 p_3 + f_2^2 \, p_3 p_1 + f_3^2 \, p_1 p_2 \right] \geq 0.
\end{equation}
Provided the principal pressures $p_i$ are all nonzero, it is sometimes useful to rewrite this as
\begin{equation}
\det\langle T^{ab}\rangle = p_1 p_2 p_3 \left\{ \rho - \sum_i { f_i^2\over p_i} \right\} \geq 0.
\end{equation}
Provided the energy density $\rho$ is nonzero, it is useful to define $w_i = p_i/\rho$, and rewrite the condition as
\begin{equation}
w_1 w_2 w_3 - \left[ \left(f_1/\rho\right)^2 \, w_2 w_3 + \left(f_2/\rho\right)^2 \, w_3 w_1 + \left(f_3/\rho\right)^2 \, w_1 w_2 \right] \geq 0.
\end{equation}
When the 3-fluxes $f_i$ are zero this simplifies to
\begin{equation}
w_1 w_2 w_3  \geq 0.
\end{equation}
It was this simple formulation in terms of a constraint on the $w_i$ parameters that first attracted our attention. 
For the four canonical types we see that
\begin{equation}
\det\langle T^{ab}\rangle:   \qquad \rho\, p_1 p_2 p_3; \qquad -\mu^2 \,p_2 p_3; \qquad \rho^3 p_3; \qquad -(\rho^2+f^2) \, p_2 p_3. 
\end{equation}
\\In the absence of other symmetries these can in principle all be of arbitrary sign.

A quantum version of the determinant bound might relax this somewhat
\begin{equation}
\det\langle T^{ab}\rangle \geq -|\hbox{some quantum bound}|.
\end{equation}
It will already be interesting if this quantity is generically bounded below, even if the specific value might be situation dependent.  This is the weak quantum determinant energy condition. If one wishes a more explicit bound to test one might formulate a strong QDETEC using a similar bound as that suggested for the QFEC. That is:
\begin{equation}
\det\langle T^{ab}\rangle \geq - (\hbar N/L^4)^4.
\end{equation}
(Thus, by definition, the QDETEC makes sense only in those situations for which there is a well-defined notion of system size.)

\section{Trace-of-square energy condition}\label{sTOSEC}

Consider the condition
\begin{equation}
\langle T^{ab}\rangle\,\langle T_{ab}\rangle \geq 0.
\end{equation}
This is essentially a Lorentzian signature analogue of the square of the Frobenius norm.
It is \emph{only} the fact that we are working in Lorentzian signature that makes this statement non-vacuous. Specifically, 
in an orthonormal frame where the spatial stress has been diagonalized, (\emph{cf} equation (\ref{E:diag})), this condition becomes
\begin{equation}
\rho^2 + \sum p_i^2 - 2 \sum f_i^2 \geq 0.
\end{equation}
In terms of the Hawking--Ellis classification of stress-energy tensors~\cite{Hawking-Ellis}, we have
\begin{equation}
\langle T^{ab}\rangle\,\langle T_{ab}\rangle:  \quad \rho^2+\sum p_i^2; \qquad 2\mu^2 + \sum p_i^2; \qquad  3\rho^2+p^2;  
\qquad 2(\rho^2-f^2) + \sum p_i^2.
\quad
\end{equation}
Thus this condition is automatically trivial for type I, type II, and type III stress tensors, but \emph{may} be nontrivial for type IV.  
(If one ever finds a situation where this trace of square energy condition is violated then the stress-energy tensor must at least locally be of type IV.)
A (weak) quantum version of the trace-squared bound might relax this somewhat
\begin{equation}
\langle T^{ab}\rangle\,\langle T_{ab}\rangle \geq -|\hbox{some quantum bound}|.
\end{equation}
Again, a more explicit (strong) formulation of the QTOSEC might be
\begin{equation}
\langle T^{ab}\rangle\,\langle T_{ab}\rangle \geq - (\hbar N/L^4)^2.
\end{equation}
(Thus, by definition, the QTOSEC makes sense only in those situations for which there is a well-defined notion of system size.)

\section{The QECs: QWEC, QSEC, and QDEC}\label{sQECs}

Before checking how the new nonlinear energy conditions behave in the  quantum regime; their quantum versions being satisfied in a rather wide range of quantum vacuum state examples, we find it interesting to formulate the quantum generalisation of the usual classical (linear) energy conditions. As we will show, this quantum version of the usual linear energy conditions is not generically satisfied. Therefore, the fulfillment of the quantum nonlinear energy conditions is not trivially due to the quantum aspects of the generalisation; something deeper is at play. We will take the WEC as the canonical example of the behaviour of the other classical (linear) energy conditions.

As is well known, the WEC states that $\langle T^{ab}\rangle \, V_a V_b \geq 0$, for any timelike vector. Therefore, taking into account~(\ref{E:diag}), the WEC can be written as:
\begin{equation}\label{WECbeta}
\rho  - 2 {\textstyle\sum_i} \beta_i f_i  + {\textstyle\sum_i} p_i\beta_i^2
\geq 0.
\end{equation}
A weak quantum version of the WEC (the QWEC) can be formulated by requiring that violations of the WEC, $\langle T^{ab}\rangle \, V_a V_b \geq 0$, should be bounded from below. That is equivalent to demanding that the energy density measured by any observer cannot be ``excessively negative". This condition can be made  more specific by formulating a strong version of the QWEC. We tentatively choose
\begin{equation}\label{E:qwec}
\langle T^{ab}\rangle \, V_a V_b \geq -\zeta(\hbar N/L^4)\; (U_a\,V^a)^2.
\end{equation}
Now, considering~(\ref{E:diag}) this can be written as
\begin{equation}\label{QWECbeta}
\rho  - 2 {\textstyle\sum_i} \beta_i f_i  + {\textstyle\sum_i} p_i\beta_i^2
\geq -\zeta (\hbar N/L^4).
\end{equation}

\paragraph{Type I.}
It is well known that for a type I tensor the WEC is equivalent to the following conditions
\begin{equation}
\rho+p_i\geq0; \qquad  \rho\geq0.
\end{equation}
Thus, in this case a strong formulation of the QWEC (which is compatible with choosing $\zeta=1$) requires
\begin{equation}
\rho+p_i\geq-(\hbar N/L^4); \qquad  \rho\geq-(\hbar N/L^4).
\end{equation}

\paragraph{Type II.} 
In this case, equation~(\ref{WECbeta}) can be written as
\begin{equation}
{\rm WEC:}\qquad
(1-\beta_1^2)\mu+(1-\beta_1)^2f+\beta_2^2p_2+\beta_3^2p_3 \geq 0,
\end{equation}
which, assuming that $\mu\geq0$, implies
\begin{equation}\label{wecs2}
\hbox{LHS}\geq\beta_2^2(\mu+p_2)+\beta_3^2(\mu+p_3)+(1-\beta_1)^2f.
\end{equation}
Therefore, a set of sufficient conditions for the WEC to be satisfied is
\begin{equation}
\mu+p_2\geq0,\qquad\mu+p_3\geq0,\qquad f\geq0,\qquad \mu\geq0,
\end{equation}
where the last condition has already been assumed at a previous stage.
This set of sufficient conditions are easier to be satisfied than those included in Hawking and Ellis~\cite{Hawking-Ellis}, since we are allowing negative $p_2$ and $p_3$ provided they are bounded by $-\mu$. 
On the other hand, the situation is more subtle in the case of the QWEC for type II, since we cannot perform the step necessary to get an equation similar to~(\ref{wecs2}). In this case, we instead have
\begin{equation}
{\rm QWEC}: \qquad
(1-\beta_1^2)\mu+(1-\beta_1)^2f+\beta_2^2p_2+\beta_3^2p_3 \geq-\zeta(\hbar N/L^4).
\end{equation}
Thus, a sufficient set of conditions for QWEC is
\begin{equation}
p_2\geq- \zeta(\hbar N/L^4),\qquad p_3\geq-\zeta(\hbar N/L^4),\qquad f\geq-\zeta(\hbar N/L^4),\qquad \mu\geq-\zeta(\hbar N/L^4).
\end{equation}

\paragraph{Type III.}
A similar argument for type III stress tensors leads to
\begin{eqnarray}
{\rm WEC:}&\quad&
(1-\beta_1^2-\beta_2^2)\rho-( \sqrt{2} \beta_1+\sqrt{2}\beta_2-\beta_1^2+\beta_2^2)f+\beta_3^2p_3 \geq 0,\qquad\qquad
\end{eqnarray}
which cannot be satisfied~\cite{Hawking-Ellis}. Regarding QWEC we have
\begin{equation}
\rho\geq-\zeta(\hbar N/L^4),\qquad p_3\geq-\zeta(\hbar N/L^4),\qquad |f|\leq\zeta(\hbar N/L^4).
\end{equation}

\paragraph{Type IV.} 
In this case, we have:
\begin{equation}
{\rm WEC:}\qquad
(1-\beta_1^2)\rho-2\beta_1 f +\beta_2^2p_2+\beta_3^2p_3\geq 0,
\end{equation}
which cannot be satisfied.
For the QWEC we would need
\begin{equation}
\rho\geq-\zeta(\hbar N/L^4),\qquad p_2\geq-\zeta(\hbar N/L^4),\qquad p_3\geq-\zeta(\hbar N/L^4),\qquad |f|\leq\zeta(\hbar N/L^4).
\end{equation}
It should be noted that a quantum version of the SEC can be formulated along similar lines, simply replace $\langle T^{ab}\rangle \to \langle T^{ab}\rangle -{1\over2} \langle T\rangle \, g^{ab}$. 
Furthermore, note that unbounded violations of the WEC would necessarily imply unbounded violations of the DEC and, therefore, violate any conceivable formulation of the QDEC.

\section{Casimir vacuum}\label{sCasimir}
For the Casimir vacuum we have
\begin{equation}
\label{E:casimir}
\langle T^{ab} \rangle =  - {\hbar \pi^2\over720 a^4} \left[\begin{array}{ccc|c}1&0&0&0\\0&-1&0&0\\0&0&-1&0\\ \hline 0&0&0&3\end{array}\right].
\end{equation}
The relevant inequalities for the FEC to be fulfilled are (marginally) satisfied in the x and y directions, (since $p_1^2/\rho^2=1= p_2^2/\rho^2$), but the inequality corresponding to the $z$ direction \emph{fails}, (since $p_3^2/\rho^2=3 > 1$). In contrast, the QFEC, (with $L\to a$, the distance between the plates), is satisfied since all the inequalities are fulfilled. The key point is that
\begin{equation}
\label{CFEC}
\rho^2 - p_3^2 = - 8 \left({\hbar \pi^2\over720 a^4}\right)^2 = - {8 \pi^4\over720^2} \left({\hbar \over a^4}\right)^2,
\end{equation}
and $8\pi^2/(720)^2 \ll 1$. 
Furthermore note
\begin{equation}
\det\langle T^{ab}\rangle = +3 \left({\hbar \pi^2\over720 a^4}\right)^4 > 0,
\qquad
\hbox{and}
\qquad
\langle T^{ab}\rangle \,\langle T_{ab}\rangle = +12 \left({\hbar \pi^2\over720 a^4}\right)^2 > 0,
\end{equation}
so both the classical determinant and trace squared conditions are certainly satisfied. The situation is summarised in table \ref{T:casimir} below.

As is well known the usual linear energy conditions are violated for the Casimir vacuum.
Nevertheless, 
taking into account equations~(\ref{E:casimir}) we see
\begin{equation}
\rho=-{\pi^2\over 720} \left({\hbar \over a^4}\right),
\qquad
\hbox{and}
\qquad
\rho+p_3=-{4\pi^2\over 720} \left({\hbar \over a^4}\right),
\end{equation}
so one can conclude that QWEC and QDEC are satisfied.
This behaviour could be viewed as suggesting that the QFEC is satisfied trivially, simply as a side effect of the QECs. 
As we will show in the following sections, that is not the case, since the QECs (though not the QFEC) are generally violated. 

\begin{table}[htdp]
\caption{Casimir spacetime.}
\begin{center}
\begin{tabular}{||c||c|c|c|c|c|c||}
\hline
\hline
Name&  usual ECs & QECs & FEC  & QFEC  & DETEC  & TOSEC \\
\hline
\hline
Status & \xmark & \cmark &  \xmark & \cmark &\cmark &\cmark \\
\hline
\hline
\end{tabular}
\end{center}
\label{T:casimir}
\end{table}%

\section{Vacuum polarization in Schwarzschild spacetime}\label{sSchw}
In this geometry the renormalized stress-energy tensor has the symmetries
\begin{equation}
\langle T^{ab}\rangle   =   
\left[\begin{array}{cc|cc}
 \rho&f&0&0\\  f & p_r & 0 & 0 \\ \hline 0 & 0 & p_t & 0 \\ 0&0&0&p_t
  \end{array}\right].
\end{equation}
Because of spherical symmetry this cannot be type III in the Hawking--Ellis classification~\cite{Hawking-Ellis}, we need only worry about types I, II, and IV. 
The renormalized stress-energy  is known to generically violate the standard energy conditions~\cite{twilight, Visser:1994, gvp1, gvp2, gvp3, gvp4, gvp5, Flanagan:1996, q-anec, volume-aec}. To see why this is plausible, there are three quantum vacuum states to consider (Boulware, Hartle--Hawking, and Unruh). 

\subsection{General analysis of the form of the stress-energy tensor}
On very general grounds,  (\emph{cf} the Page approximation~\cite{Page}, see also~\cite{gvp1,gvp2}, and~\cite{Prado:2013}), in the Hartle--Hawking vacuum we have:
\begin{equation}
\langle T^{ab} \rangle_{HH} = \O(1).
\end{equation}
The key point is that the Hartle--Hawking vacuum is specifically constructed to make the renormalized stress-energy components bounded everywhere. 
Furthermore, setting $z=2M/r$, (\emph{cf} the Brown--Ottewill approximation~\cite{Brown-Ottewill}, see also~\cite{gvp1}, and~\cite{Prado:2013}), in the Boulware vacuum we have:
\begin{equation}
\langle T^{ab} \rangle_{B} = -{p_\infty\over(1-z)^2}   
\left[\begin{array}{cc|cc}
 3&0&0&0\\  0 & 1 & 0 & 0 \\ \hline 0 & 0 & 1 & 0 \\ 0&0&0&1
  \end{array}\right] + 
\langle T^{ab} \rangle_{HH}  +\O(1)
=
 -{p_\infty\over(1-z)^2}   
\left[\begin{array}{cc|cc}
 3&0&0&0\\  0 & 1 & 0 & 0 \\ \hline 0 & 0 & 1 & 0 \\ 0&0&0&1
  \end{array}\right] 
+\O(1),
\end{equation}
with $p_\infty$ a positive constant depending on the specific QFT under consideration.
The key point is that in the Boulware vacuum the near-horizon behaviour is dominated by a vacuum polarization term taking the form (but not the sign) of a divergent thermal bath. 
Finally, for the Unruh vacuum, based on the covariant conservation of stress-energy we can write (see for example~\cite{gvp4})
\begin{eqnarray}
\langle T^{ab} \rangle_{U} &=& {f_0 z^2 \over1-z}   
\left[\begin{array}{cc|cc}
 -1&1&0&0\\  1 & -1 & 0 & 0 \\ \hline 0 & 0 & 0 & 0 \\ 0&0&0&0
  \end{array}\right] 
 + \left[\begin{array}{cc|cc}
 -T(z)+2p_t(z)+ K(z)&0&0&0\\  0 & K(z) & 0 & 0 \\ \hline 0 & 0 & p_{t}(z) & 0 \\ 0&0&0&p_{t}(z)
  \end{array}\right].\qquad
\end{eqnarray}
Here
\begin{equation}
K(z) = {z^2\over1-z} [H(z)+G(z)],
\end{equation}
where
\begin{equation}
H(z) = {1\over2} \int_z^1 {T(z)\over z^2} \d z; \qquad G(z) = \int_z^1 \left({2\over z^3} - {3\over z^2} \right) p_t(z) \d z.
\end{equation}
Let $p_H$ denote the on-horizon transverse pressure, $T_H$ denote the on-horizon trace of the stress-energy (given by the conformal anomaly), then 
\begin{equation}
H(z) = {1\over2} T_H (1-z) + \O([1-z]^2); \qquad G(z) = - p_H (1-z) + \O([1-z]^2),
\end{equation}
so that
\begin{equation}
K(z) = {1\over2} T_H - p_H + \O(1-z).
\end{equation}
Finally defining $\Delta = {T_H\over2}-p_H$ one has:
\begin{eqnarray}
\langle T^{ab} \rangle_{U} &=& {f_0 z^2 \over1-z}   
\left[\begin{array}{cc|cc}
 -1&1&0&0\\  1 & -1 & 0 & 0 \\ \hline 0 & 0 & 0 & 0 \\ 0&0&0&0
  \end{array}\right] 
 + \left[\begin{array}{cc|cc}
 -\Delta&0&0&0\\  0 & \Delta & 0 & 0 \\ \hline 0 & 0 & p_{H} & 0 \\ 0&0&0&p_{H}
  \end{array}\right] 
+\O(1-z).
\end{eqnarray}

\subsection{Implications for the energy conditions}

Sufficiently close to the horizon the pole pieces in the stress-energy dominate, and violation of the usual energy conditions is automatic for both the Boulware and Unruh vacuum states. For the Hartle--Hawking state we merely know that the stress-energy components are bounded, and to verify violation of the usual energy conditions one has to resort either to direct inspection of the numerical data~\cite{gvp1}, or work with known analytical models able to describe such data accurately for specific QFTs.  

In the first place, we include some results regarding the FEC/QFEC which have already been outlined in~\cite{Prado:2013}. The relevant quantities to consider in the Hartle--Hawking and Boulware vacua are
\begin{equation}
\{ \rho^2-p_i^2 \}_{HH} = \O(1),
\end{equation}
and
\begin{equation}
\{ \rho^2-p_i^2 \}_{B} = {8p_\infty^2\over(1-z)^4} +\O([1-z]^{-2}).
\end{equation}
As both stress energy tensors are type I, conditions~(\ref{E:seven}) are necessary and sufficient conditions for the FEC to be satisfied.
Thus, we can conclude that the violations of the FEC, if any, are bounded from below  for the Hartle--Hawking and Boulware vacua. 
This is enough to guarantee the weak form of the QFEC is satisfied.

On the other hand, we cannot extract any definitive conclusion regarding the Unruh vacuum without restricting attention to a particular QFT. This is due to the fact that one needs to know what type of stress energy tensor it is to consider the particular \emph{sufficient constraints} for that case. All we can do in general is to consider \emph{necessary} constraints, as those given by~(\ref{E:seven}). The relevant quantities appearing in those constraints are
\begin{equation}
\{ \rho^2-f^2-p_t^2 \}_{U} ={2f_0\Delta\over1-z}+\O(1);
\end{equation}
\begin{eqnarray}
\{ (\rho+f)^2-(p_r+f)^2 \}_{U} &=& \{ (\rho-p_r)(\rho+p_r+2f) \}_{U} = \O(1-z);
\end{eqnarray}
\begin{eqnarray}
\{ (\rho-f)^2-(p_r-f)^2 \}_{U} &=& \{ (\rho-p_r)(\rho+p_r-2f) \}_{U} = {8f_0\Delta\over1-z}+\O(1).
\end{eqnarray}
Thus, whereas for $\Delta< 0$ we can conclude that the violations of the FEC are unbounded, since they indeed are for some particular observers, for $\Delta\geq0$ further considerations are needed restricting attention to the particular model.

In the second place, for the determinant energy condition the relevant quantities to consider in the Hartle--Hawking and Unruh vacua are
\begin{equation}
\det\langle T^{ab}\rangle_{HH} = \O(1),
\end{equation}
and
\begin{equation}
\det\langle T^{ab}\rangle_{B} = {3p_\infty^4\over(1-z)^8} +\O([1-z]^{-6}).
\end{equation}
More subtly in the Unruh vacuum there are significant cancellations in calculating the determinant and one should consider 
\begin{equation}
\det\langle T^{ab}\rangle_{U} = \left\{ -\Delta^2 + f_0\times \O(1) \right\} p_H^2 = \O(1).  
\end{equation}
The determinant is certainly bounded from below in all three vacua, and is even bounded both above and below in the Hartle--Hawking and Unruh vacua. 

In the third place, for the trace-of-square condition the relevant quantities to consider in the Hartle--Hawking and Unruh vacua are
\begin{equation}
\left(\langle T^{ab}\rangle \,\langle T_{ab}\rangle\right)_{HH} = \O(1),
\end{equation}
and
\begin{equation}
\left(\langle T^{ab}\rangle \,\langle T_{ab}\rangle\right)_{B} = +{12p_\infty^2\over(1-z)^4} +\O([1-z]^{-2}).
\end{equation}
In fact because these two stress-energy tensors are diagonal both these quantities are actually guaranteed non-negative. 
More subtly in the Unruh vacuum there are again significant cancellations in the calculation and one should consider 
\begin{equation}
\left(\langle T^{ab}\rangle \,\langle T_{ab}\rangle\right)_{U} = 2 \Delta^2 +4 f_0\times \O(1) + 2\,p_H^2 = \O(1). 
\end{equation}
The trace-squared is certainly bounded from below in all three vacua, and is even bounded both above and below in the Hartle--Hawking and Unruh vacua. 

Finally, it can be trivially seen that close to the horizon the violations of the standard energy conditions are unbounded for the Boulware and Unruh vacuum. We include for completeness the relevant quantities to consider for studying the QWEC. We see
\begin{equation}
\{ \rho+p_i \}_{HH} = \O(1),\qquad \{ \rho\}_{HH} = \O(1),
\end{equation}
and
\begin{equation}
\{ \rho+p_i \}_{B} = -{4 p_\infty \over(1-z)^2} +\O(1)\qquad \{ \rho\}_{B} = -{3 p_\infty \over(1-z)^2} +\O(1).
\end{equation}
The Hartle--Hawking vacuum is particularly well-behaved, since even the violations of the standard energy conditions are bounded, allowing the fulfillment of weak formulation of the quantum version. For the Boulware vacuum, the QWEC is violated, since $p_\infty>0$.
For the Unruh vacuum the three quantities used to formulate  the necessary conditions are
\begin{equation}
\{ \rho+p_r+2f \}_{U} =\O([1-z]);
\qquad
\{ \rho+p_r-2f \}_{U} =-{4 f_0 z^2\over 1-z} +\O([1-z]);
\end{equation}
and
\begin{equation}
\qquad
\{ \rho+p_t\}_{U} =-{ f_0 z^2\over 1-z} +\O(1).
\end{equation}
As $f_0>0$, the necessary conditions are violated in an unbounded way, suggesting that the QWEC is violated.
The overall generic situation is summarised in table~\ref{T:generic}.

\begin{table}[!htdp]
\caption{Schwarzschild geometry: Generic situation.}
\begin{center}
\begin{tabular}{||c||c|c|c|c||}
\hline
\hline
Name                  &  usual ECs & FEC  & DETEC & TOSEC \\
\hline
\hline
Hartle--Hawking          & bounded  & bounded & bounded &\cmark \\
\hline
Boulware  & \xmark  & bounded &bounded &\cmark \\
\hline
Unruh                & \xmark &  ? & bounded & bounded \\
\hline
\hline
\end{tabular}
\end{center}
\label{T:generic}
\end{table}%

\subsection{Specific example: Massless conformally coupled scalar}

Let us now study the specific case of a massless conformally coupled scalar field in Schwarzschild spacetime. 

\subsubsection{Hartle--Hawking vacuum}

For the Hartle--Hawking vacuum one has a useful analytic approximation to the gravitational polarisation, Page's approximation~\cite{Page}. (See~\cite{gvp1} for detailed discussion about the ability of this approximation to match the numerical data.) Page's approximation gives a polynomial approximation to the non-vanishing quantities appearing in a stress energy tensor of the form given in~(\ref{E:diag}) for this case. These are~\cite{gvp1}:
\begin{equation}\label{rH}
\{\rho\}_{HH}(z)=3\,p_\infty\left[1+2\,z+3\,z^2+4\,z^3+5\,z^4+6\,z^5-33\,z^6\right],
\end{equation}
\begin{equation}\label{p1H}
\{p_r\}_{HH}(z)=p_\infty\left[1+2\,z+3\,z^2+4\,z^3+5\,z^4+6\,z^5+15\,z^6\right],
\end{equation}
and
\begin{equation}\label{p2H}
\{p_{t}\}_{HH}(z)=p_\infty\left[1+2\,z+3\,z^2+4\,z^3+5\,z^4+6\,z^5-9\,z^6\right],
\end{equation}
where
\begin{equation}
p_\infty={\hslash\over 90(16\pi)^2(2M)^4}.
\end{equation}
As has already been seen in~\cite{gvp1} the NEC is violated for this vacuum because, although $\{\rho+p_r\}_{HH}>0$, we have $\{\rho\}_{HH}<0$ and
 $\{\rho+p_{t}\}_{HH}<0$ close to the horizon. Nevertheless, this violation is bounded from below. It can be easily seen that the largest violation would be at the horizon where
\begin{equation}
\{\rho\}_{HH}=-36\,p_\infty>-{\hslash\over(2M)^4},\qquad\{\rho+p_{t}\}_{HH}=-24\,p_\infty>-{\hslash\over(2M)^4},
\end{equation}
therefore, the QWEC is fulfilled.

\begin{figure}[!htbp]
\begin{center}
\includegraphics[scale=0.75]{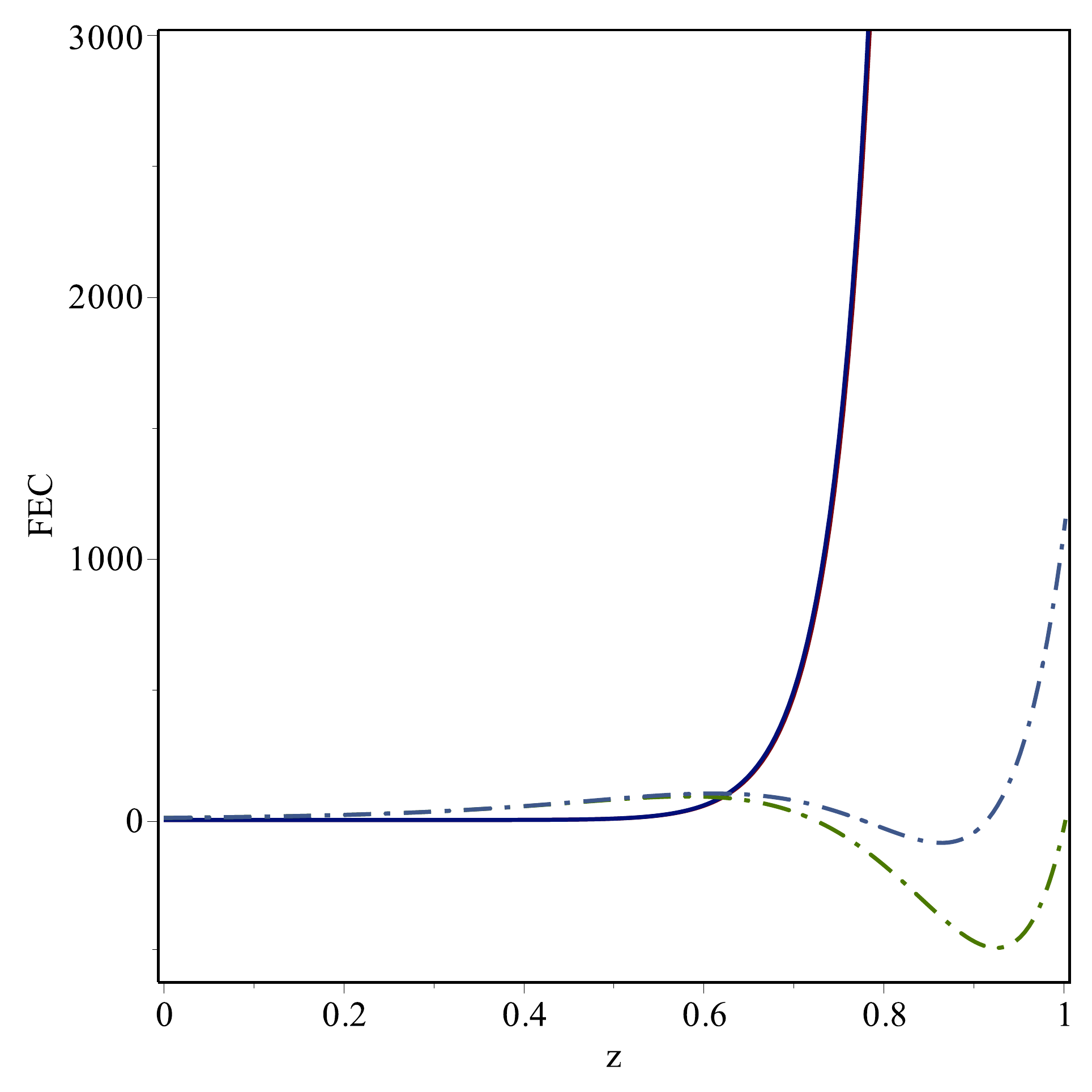}
\caption{FEC for the conformally coupled massless scalar field.\newline
Various FEC-related quantities are plotted for the Boulware (solid) and Hartle--Hawking (dashed) quantum states. 
When these quantities go negative it indicates the FEC is violated. The boundedness of the negative excursion indicates the weak QFEC is satisfied. 
We use units such that $p_\infty=1$.
}
\label{F:fec}
\end{center}
\end{figure}

Again taking into account~(\ref{rH}), (\ref{p1H}), and (\ref{p2H}), one can conclude that the FEC is violated close to the horizon ($r\lesssim2.7725M$), where first $\{\rho^2-p_r^2\}_{HH}$ and then $\{\rho^2-p_{t}^2\}_{HH}$ take negative values. Considering the largest violation in one of the two conditions (at $r\sim 2.1663M$) one can see that
\begin{equation}
\{\rho^2-p_1^2\}_{HH}>-489.2897\, p_\infty^2>-{\hslash^2\over(2M)^8}.
\end{equation} 
Thus, the strong QFEC is fulfilled. (See figure~\ref{F:fec}.)

On the other hand, it can be seen that the determinant condition is also violated. Such violation is, however, bounded from below, with the largest violation occurring at the horizon where
\begin{equation}
{\rm det}\langle T^{ab}\rangle\geq \{\rho\,p_r\,p_t^2\}_{HH}(z=1) =-186624\, p_\infty^4>-{\hslash^4 \over (2M)^{16}}.
\end{equation}
 (See figure~\ref{F:det}.) Therefore, a quantum version of the DETEC is satisfied.

Finally, it must be noted that as this stress energy tensor is type I, the TOSEC is trivially satisfied.

\subsubsection{Boulware vacuum}

A simple polynomial approximation to the non-vanishing quantities appearing in a stress energy tensor of the form given in~(\ref{E:diag}) can be obtained for the Boulware vacuum~\cite{gvp2}, by combining Page's approximation~\cite{Page} with the results of Brown and Ottewill~\cite{Brown-Ottewill}. These are~\cite{gvp2}:
\begin{equation}
\label{rB}
\{\rho\}_B(z)=-3\,p_\infty \, z^6\;{40-72\,z+33\,z^2\over \left(1-z\right)^2},
\end{equation}
\begin{equation}
\label{p1B}
\{p_r\}_B(z)=p_\infty \,z^6\;{8-24\,z+15\,z^2\over \left(1-z\right)^2},
\end{equation}
and
\begin{equation}
\label{p2B}
\{p_{t}\}_B(z)=-p_\infty \,z^6\;{\left(4-3\,z\right)^2 \over \left(1-z\right)^2}.
\end{equation}
As we have already pointed out in the preceding subsection, the violations of the classical linear energy conditions are unbounded for this vacuum. Thus, a quantum version of the usual energy conditions cannot be satisfied.

Let us now explicitly write the LHS of the inequalities which should be fulfilled for the FEC to hold (since for this vacuum they are particularly simple). These are:
\begin{equation}
\{\rho^2-p_r^2\}_{B}=8\,p_\infty^2 \,z^{12}\; {(64-120z+57z^2)(28-48z+21z^2)\over \left(1-z\right)^4},
\end{equation} 
and
\begin{equation}
\{\rho^2-p_t^2\}_{B}=8\,p_\infty^2 \, z^{12}\; {(52-96z+45z^2)(34-60z+27z^2)\over \left(1-z\right)^4}.
\end{equation} 
Both quantities are positive for any $0<z<1$. Thus, the FEC (and hence the QFEC) is fulfilled.
Regarding the TOSEC and the DETEC we have a behaviour similar than that of the Hartle--Hawking vacuum. That is, the first is trivially fulfilled. Violation of the DETEC is limited to the range $z \lesssim 0.47340$, and is again bounded. In fact the bound is numerically rather small, ${\rm det}(T^{ab})>-0.0003\,p^4_\infty$, with the minimum occurring near $z\approx 0.45624$. The smallness of this numerical bound is largely due to a factor of $z^{24}$ in the determinant.

\subsubsection{Unruh vacuum}

For this vacuum no true analytic approximation is available. However, a semi-analytical approximation has been presented in~\cite{gvp4},  obtained by data-fitting to a semi-empirical model.  The relevant non-vanishing quantities in that semi-analytical model are:
\begin{equation}
\label{rU}
\{\rho\}_U=p_\infty\, {z^2\over 1-z}\;\left(5.349+26.56\,z^2-105.2\,z^3+35.09\,z^4+32.91\,z^5\right),
\end{equation}
\begin{equation}
\label{p1U}
\{p_r\}_U=p_\infty\, {z^2\over 1-z}\;\left(5.349-26.56\,z^2+65.92z^3-63.37\,z^4+13.32\,z^5\right),
\end{equation}
\begin{equation}
\label{p2U}
\{p_{t}\}_U=p_\infty\, z^4\;\left(26.56-59.02\,z+38.21\,z^2\right),
\end{equation}
and
\begin{equation}\label{fU}
\{f\}_U=5.349\,p_\infty \; {z^2\over 1-z}.
\end{equation}
To consider the QFEC, in the first place, we must study of which type the stress energy tensor is. Thus, taking $\Gamma=(\rho+p_r)^2-4f^2$, one has
\begin{eqnarray}
&&\Gamma=\frac{z^7p_\infty^2}{(1-z)^2}
\\
&&\qquad \times\left(-842.8-605.1 z+9891 z^2+1552 z^3+2228 z^4-2842z^5-2615z^6+2137z^7\right).
\nonumber
\end{eqnarray}
Thus, close to the horizon ($z\sim1$), one has 
\begin{equation}
\Gamma\sim0.9434\;\frac{p_\infty^2}{(1-z)^2}>0,
\end{equation}
and the Unruh vacuum is type I; whereas in the asymptotic region ($z\sim0$) we have
\begin{equation}
\Gamma\sim-842.8\;\frac{z^7p_\infty^2}{(1-z)^2}<0,
\end{equation}
being type IV. Moreover, one can study the quantity $\Gamma$ to conclude that it only changes sign once in the interval, at $z\sim0.9843$. Therefore, the stress energy tensor is type IV for $z\aplt0.9843$, and type I just close to the horizon ($z\apgt0.9843$).

\begin{figure}[!htbp]
\begin{center}
\includegraphics[scale=0.75]{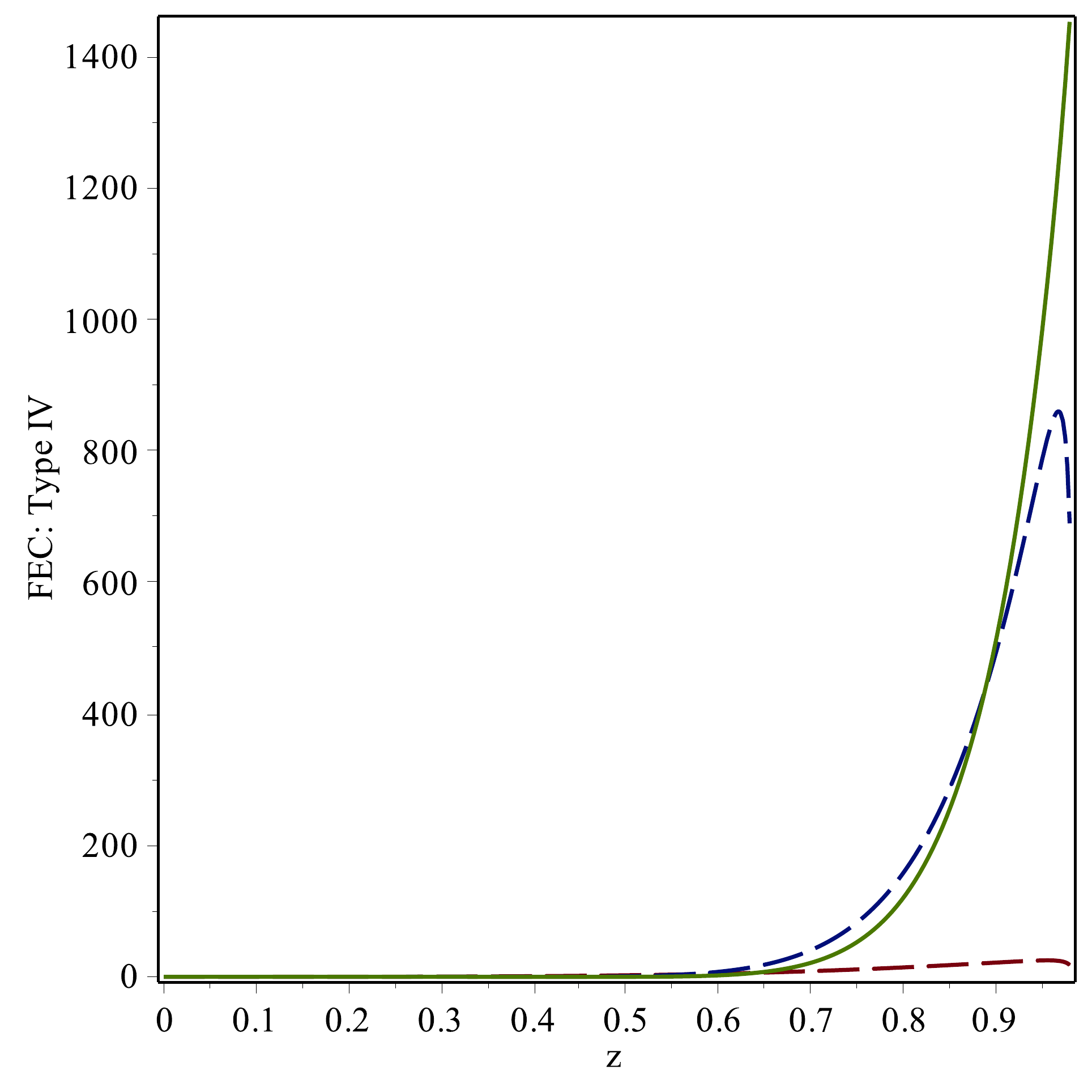}
\caption{FEC for the conformally coupled massless scalar field for the Unruh vacuum in the region where the stress energy tensor is type IV. \newline
The dashed lines represent quantities which must be bounded from above, whereas the continuous line must be bounded from below. 
Thus, the QFEC is satisfied for the Unruh vacuum in this region. }
\label{F:UtypeIV}
\end{center}
\end{figure}

In the second place, let us consider the type IV region. A type IV stress energy tensor cannot fulfill the FEC. Nevertheless, if the violations are bounded it can at least satisfied the strong formulation of the QFEC. To study this, we must note that the stress energy tensor is not written in the canonical form (\ref{qfecIV}). Therefore, for this study we need to calculate the eigenvalues of the mixed tensor $T^a{}_b$. Two of them are equal to $p_t$, whereas the other two are complex with
\begin{equation}
{\rm Re}\left(\lambda\right)={\rm Re}\left(\bar\lambda\right)=-\frac{\rho-p_r}{2},\qquad
{\rm Im}\left(\lambda\right)=-{\rm Im}\left(\bar\lambda\right)=-\frac{\sqrt{-\Gamma}}{2}.
\end{equation}
The fulfilment of the QFEC requires, on one hand, that
\begin{equation}
|{\rm Im}(\lambda)|\leq \zeta\hbar N/L^4,\qquad  |{\rm Re}(\lambda){\rm Im}(\lambda)|\leq \zeta (\hbar N/L^4)^2.
\end{equation}
It can be seen that the absolute value of these quantities is bounded, see figure~\ref{F:UtypeIV}. Indeed the greatest value taken by one of these quantities is $<900p_\infty\ll\hbar/(2M^4)$. On the other hand, we must consider
\begin{equation}
 \left({\rm Re}(\lambda)\right)^2-p_{t}^2\geq-\zeta(\hbar N/L^4)^2.
\end{equation}
One can calculate that this quantity is bounded from below by $-0.0991\,p_\infty^2\gg-\hbar/(2M^4)$. Therefore, the QFEC is fulfilled in the region where the stress energy tensor is type IV.

\begin{figure}[!htbp]
\begin{center}
\includegraphics[scale=0.75]{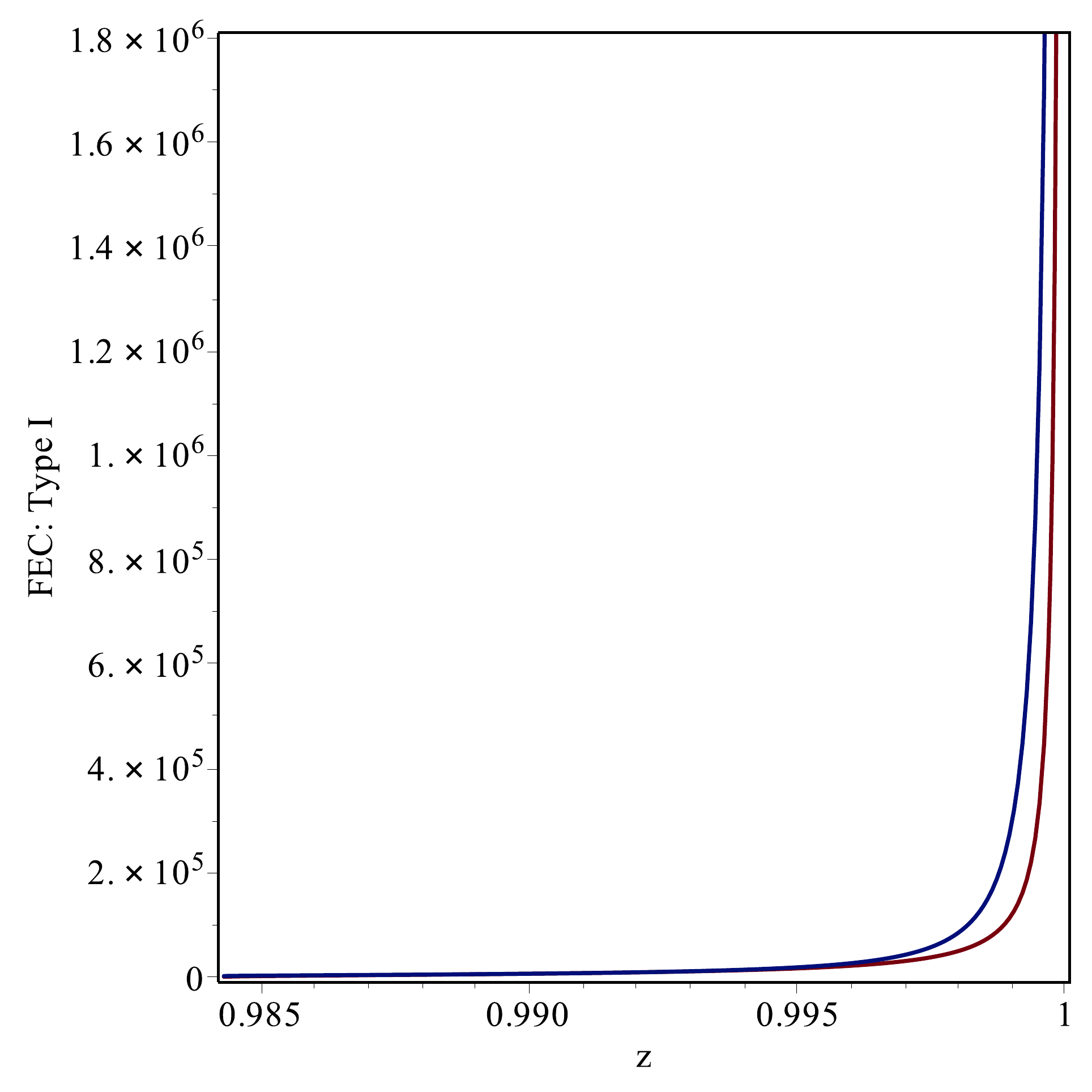}
\caption{FEC for the conformally coupled massless scalar field for the Unruh vacuum in the region where the stress energy tensor is type I. \newline
The relevant quantities are larger than zero; therefore the FEC is satisfied close to the horizon. }
\label{UtypeI}
\end{center}
\end{figure}

In the third place, we study the region where the stress energy tensor is type I. In this region the eigenvalues are
\begin{equation}
\lambda_0=-\frac{\rho-p_r}{2}+\frac{\sqrt{\Gamma}}{2},\qquad \lambda_1=-\frac{\rho-p_r}{2}-\frac{\sqrt{\Gamma}}{2},
\end{equation}
and, of course,  the double eigenvalue $p_t$. Studying the eigenvectors associated with these eigenvalues, one can conclude that $\lambda_0$ is associated to a timelike eigenvector, whereas the eigenvector associated with $\lambda_1$ is spacelike. Thus, the relevant quantities to be considered are
\begin{equation}
\lambda_0^2-\lambda_1^2,\qquad \lambda_0^2-p_t^2.
\end{equation}
In figure~\ref{UtypeI} we show that these quantities are in fact positive in the interval.
Finally, it can be noted that at the point where the stress energy tensor is type II, it is expressed as in~(\ref{E:HE2b}) with $\{f\}_U=\bar f$, and the constraints~(\ref{cfec3b}), $\mu \bar f<0$ and $\lambda^2-p_t^2>0$, are satisfied.
In summary, since the QFEC is satisfied in any region outside the horizon, the QFEC is fulfilled for the Unruh vacuum in 1+3-dimensional Schwarzschild spacetime.

\begin{figure}[!htbp]
\begin{center}
\includegraphics[scale=0.75]{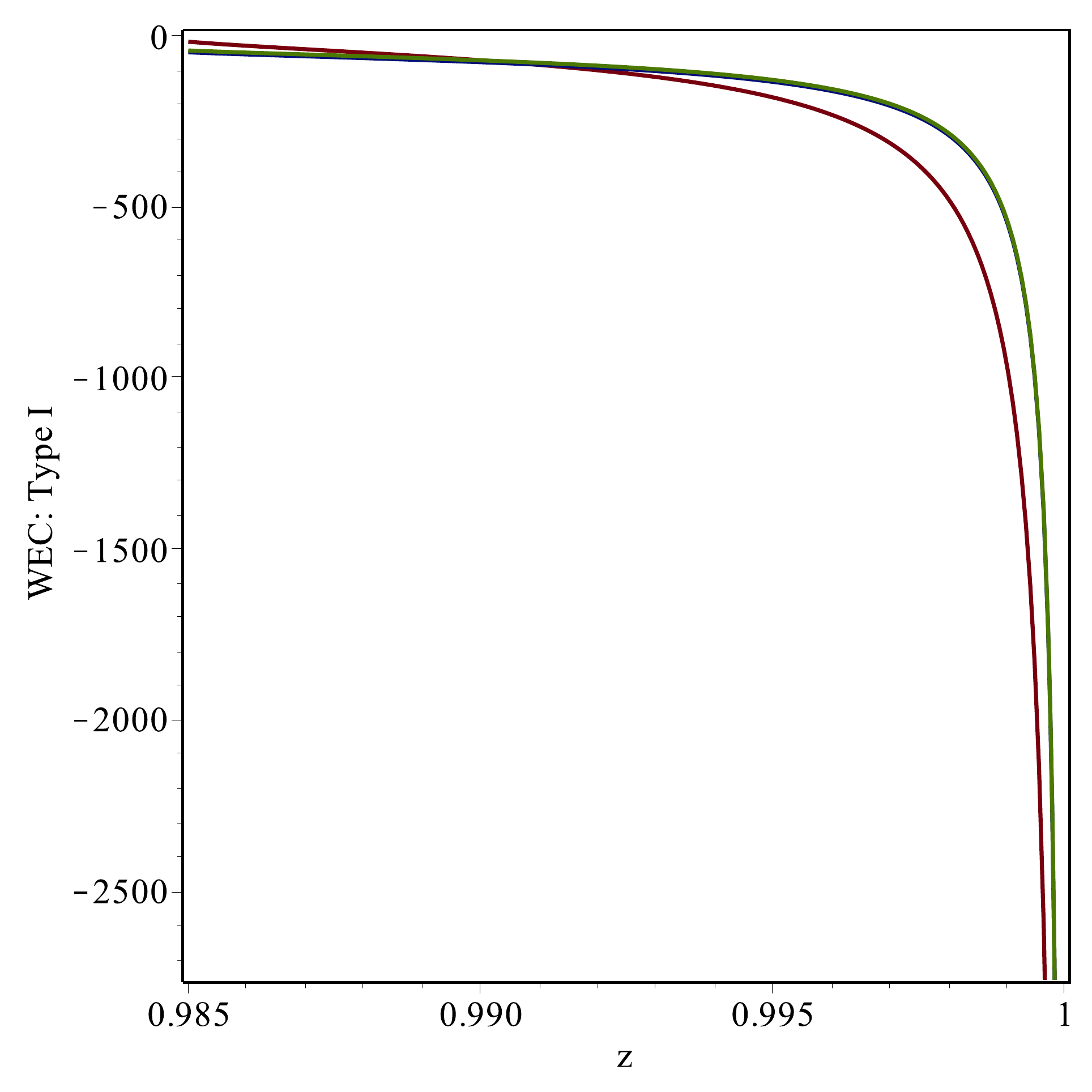}
\caption{WEC for the conformally coupled massless scalar field for the Unruh vacuum in the region where the stress energy tensor is type I.
As can be seen the QWEC is violated close to the horizon, due to unbounded violations of the WEC. }
\label{wecU}
\end{center}
\end{figure}

As is known, the WEC is violated in this case, the violation being unbounded at the horizon. Although in order to check this it is enough to consider the necessary conditions as been done in~\cite{gvp4}, here we will consider the sufficient conditions. Thus, the relevant quantities which must be bounded from below close to the horizon are
\begin{equation}
-\lambda_0,\qquad -\lambda_0+\lambda_1,\qquad-\lambda_0+p_t.
\end{equation}
As it is shown in figure~\ref{wecU}, all these quantities tend to minus infinity, demonstrating the unboundedness of the violations of the WEC.

\begin{figure}[!htbp]
\begin{center}
\includegraphics[scale=0.75]{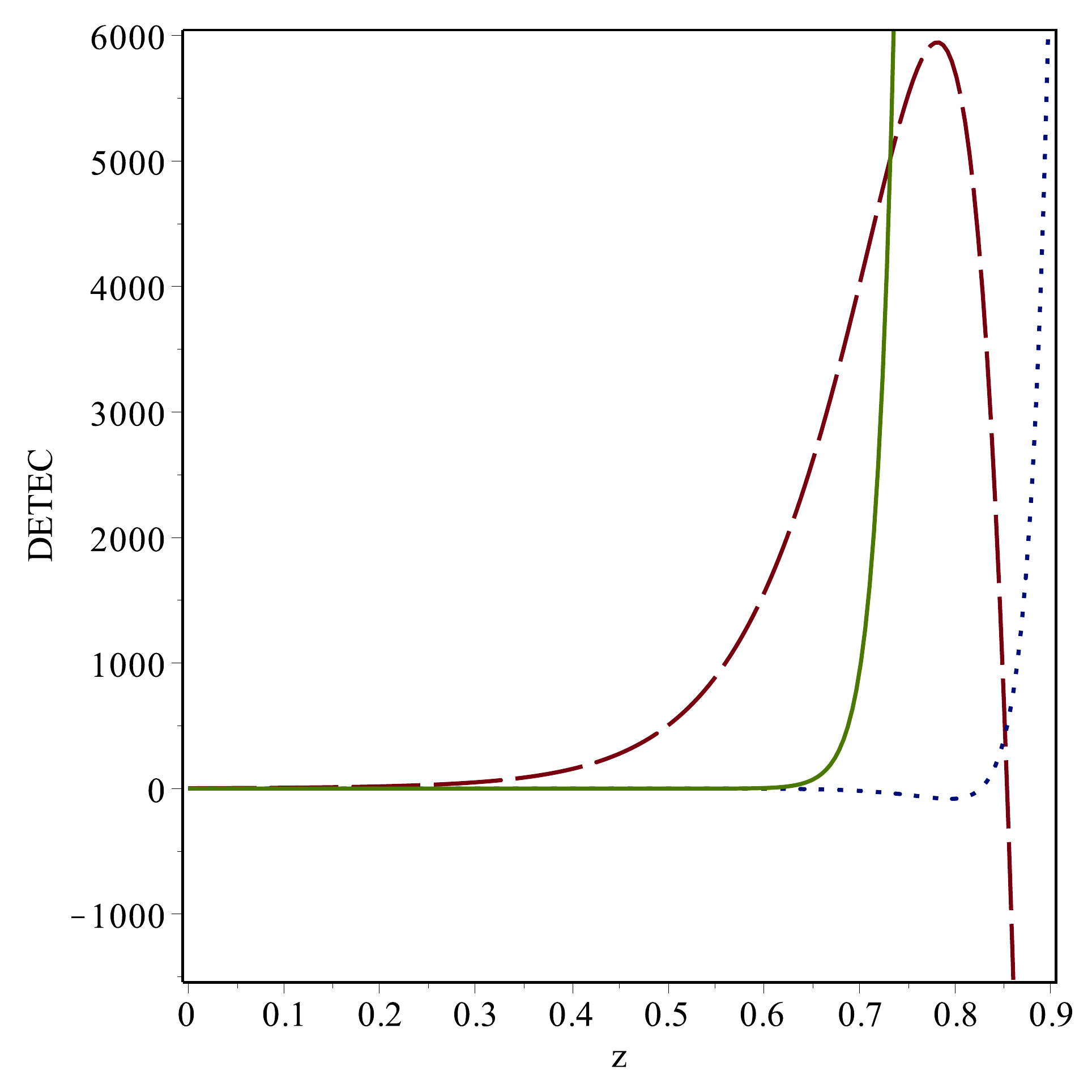}
\caption{$\det(T^{ab})$ for the conformally coupled massless scalar field. \newline
The determinant is plotted for the Boulware (solid), Hartle--Hawking (dashed), and Unruh (dotted) quantum states. 
When these quantities go negative it indicates the DETEC is violated. The boundedness of the negative excursion indicates the weak quantum DETEC is satisfied. (As indicated in the text the curve for Hartle--Hawking state is in fact bounded below by $ -186624\, p_\infty^4$, well outside the scale of the graph.) }
\label{F:det}
\end{center}
\end{figure}

\begin{figure}[!htbp]
\begin{center}
\includegraphics[scale=0.75]{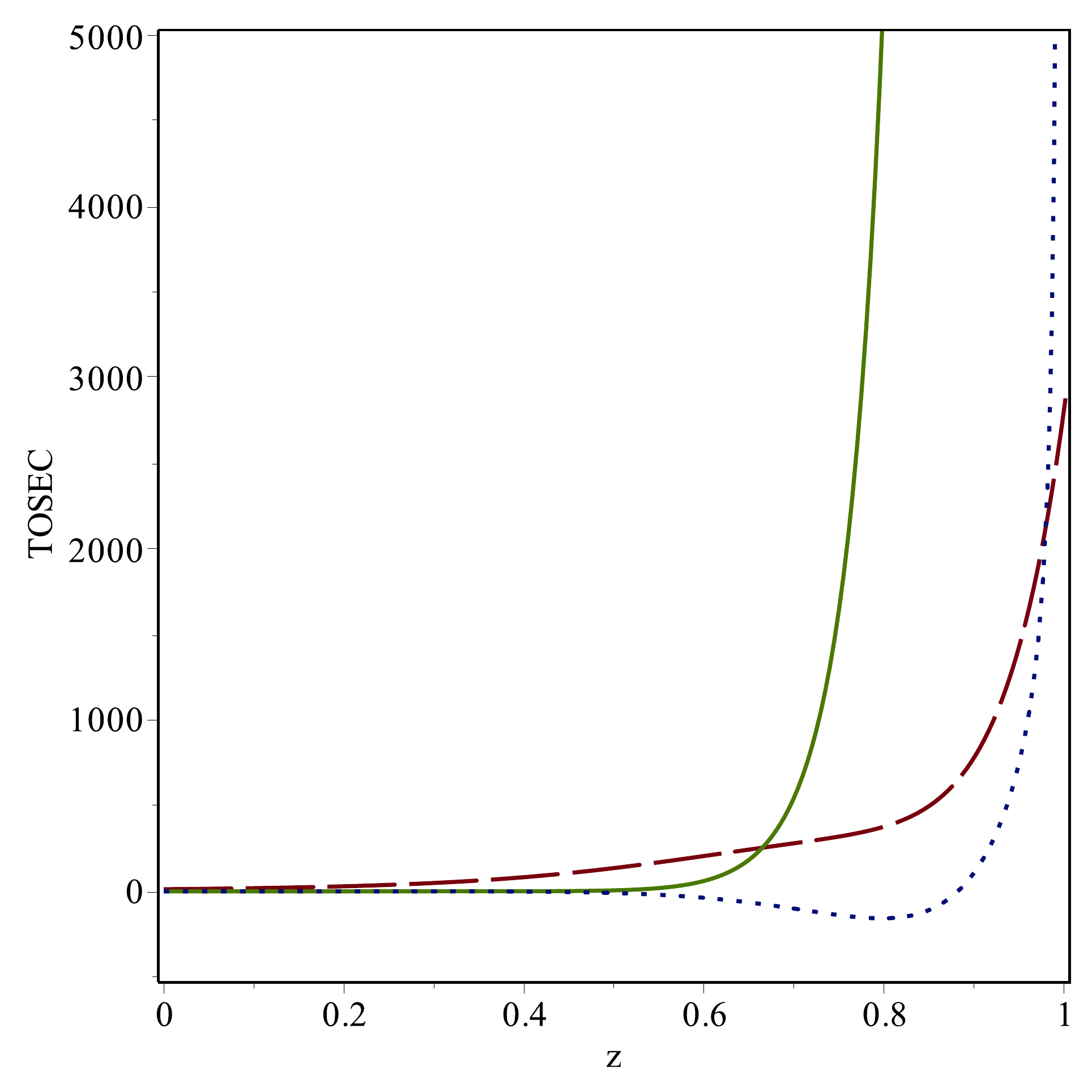}
\caption{tr$(T^2)$ for the conformally coupled massless scalar field.\newline
The trace of $T^2$ is plotted for the Boulware (solid), Hartle--Hawking (dashed), and Unruh (dotted) quantum states. 
When these quantities go negative it indicates the TOSEC is violated. The boundedness of the negative excursion indicates the weak quantum TOSEC is satisfied. 
 }
\label{F:tsq}
\end{center}
\end{figure}

On the other hand, for the Unruh vacuum not only the determinant condition but also the trace-of-square condition is violated, the violations being bounded. 
In fact, the maximum negative value takes place near $z\approx0.7959$ where ${\rm tr}(T^2)\geq-158.1\,p_\infty^2$.
(See figures~\ref{F:det} and \ref{F:tsq}.) It must be noted that this negative behaviour of the trace-of-square is possible only because the Unruh vacuum becomes of type IV in this region. 

It must be noted that when referring to the fulfillment or violation of the energy conditions we are not distinguishing (nor do we need to distinguish) between future horizon, past horizon, or bifurcation surface. Although the Unruh stress energy tensor is regular on the future horizon but singular on the past horizon, point-wise energy conditions are inequalities to be satisfied as seen by any timelike observer and, therefore, the fulfillment of these inequalities by any specific observer (whether free falling, static, or something else) is a necessary but not sufficient condition for the energy conditions to be satisfied.

The overall situation is summarized in table~\ref{T:scalar}. 

\enlargethispage{20pt}
\begin{table}[!h]
\caption{Schwarzschild geometry: Massless conformally coupled scalar.}
\begin{center}
\begin{tabular}{||c||c|c|c|c|c|c|c|c||}
\hline
\hline
Name                  &  ECs &  QECs & FEC  & QFEC  & DETEC & QDETEC& TOSEC & QTOSEC \\
\hline
\hline
H--H            & \xmark &  \cmark & \xmark & \cmark &\xmark &\cmark & \cmark&\cmark\\
\hline
Boulware 			& \xmark & \xmark & \cmark & \cmark &\xmark &\cmark &\cmark&\cmark\\
\hline
Unruh                & \xmark & \xmark &\xmark & \cmark &\xmark & \cmark &\xmark&\cmark\\
\hline
\hline
\end{tabular}
\end{center}
\label{T:scalar}
\end{table}%

\clearpage

\clearpage
\section{Vacuum polarization in 1+1 QFTs}\label{s11}
For a static geometry in 1+1 dimensions we can, without any loss of generality, set
\begin{equation}
\d s^2 = -(1-b(x))\;\d t^2 + {\d x^2\over1-b(x)}.
\end{equation}
The Ricci scalar is then
\begin{equation}
R(x) = b''(x).
\end{equation}
Furthermore, let us assume a horizon at $x_H$, so that $b(x_H)=1$, and asymptotic flatness so that $b(\infty)=0$. 
The surface gravity is then
\begin{equation}
\kappa = - {1\over2} b'(r_H).
\end{equation}

\subsection{General analysis of the form of the stress-energy tensor}

It is a standard result that for massless fields the conformal anomaly yields
\begin{equation}
\langle T\rangle = a_1 R; \qquad a_1 = {n_b+{1\over2} n_f\over 24\pi} \geq 0.
\end{equation}
This can be fed back into the equation for the covariant conservation of stress-energy to determine all the components of the stress energy up to two constants of integration --- choosing specific constants of integration is tantamount to choosing a specific quantum vacuum state. In the coordinate basis we write
\begin{equation}
\langle T^{ab}\rangle =  \left[\begin{array}{cc}\displaystyle{\rho(x)\over1-b(x)} & f(x) \\ f(x) & p(x)\; (1-b(x)) \end{array}\right],
\end{equation}
where $\rho$, $p$, and $f$ are the physical orthonormal components of the stress-energy. In an orthonormal frame
\begin{equation}
\langle T^{\hat a\hat b}\rangle =  \left[\begin{array}{cc}\rho(x) & f(x) \\ f(x) & p(x)\end{array}\right].
\end{equation}
Because of the reduced dimensionality, stress energy tensors in 1+1 dimensions can only be of type I, II, and IV in the Hawking--Ellis classification~\cite{Hawking-Ellis} --- there is no analogue of type III. 
Combining covariant  conservation of stress-energy with the conformal anomaly leads to
\begin{equation}
f = {f_0\over1-b};
\qquad\qquad
p  = \rho + a_1 R = {C\over1-b} - {a_1 (b')^2\over 4(1-b)}.
\end{equation}
The appropriate boundary conditions are:
\begin{itemize}
\item Boulware: $C=0$ and $f_0=0$.
\item Hartle--Hawking: $C=a_1 \kappa^2$ and $f_0=0$.
\item Unruh: $C = f_0 = {1\over2} a_1 \kappa^2$.
\end{itemize}

\subsection{Boulware vacuum}
For the Boulware vacuum we find:
\begin{eqnarray}
\rho_B &=& -a_1 b'' - { {1\over4} a_1 (b')^2\over1-b};
\\
p_B &=& - { {1\over4} a_1 (b')^2\over1-b}.
\end{eqnarray}
So near the horizon we have
\begin{eqnarray}
\rho_B &=& -{a_1\kappa\over2(x-x_H)} -{5\over8} a_1 R_H +\O(x-x_H);  
\\
p_B &=& - {a_1\kappa\over2(x-x_H)} +{3\over 8} a_1 R_H +\O(x-x_H). 
\end{eqnarray}
Thus, the classical linear energy conditions are all violated, the violation being unbounded at the horizon.
Now consider
\begin{equation}
\rho_{B}^2 - p_{B}^2 = a_1^2 \,b'' \, \left[  b''  + {{1\over2}  (b')^2\over1-b}\right].
\end{equation}
As has been shown in~\cite{Prado:2013}, the FEC (and hence the QFEC) is then automatically satisfied if $b'' \geq0$, that is $R\geq 0$. (These conditions are also satisfied \emph{locally} if $R \leq -{1\over2} (b')^2/(1-b)$, but applied \emph{globally} this would require infinite curvature on the horizon. This is grossly unphysical.) Near the horizon
\begin{equation}
\rho_{B}^2 - p_{B}^2 = {a_1^2\, R_H \, \kappa\over x - x_H} + \O(1).
\end{equation}
Thus $R_H \geq 0$ is necessary for the QFEC to hold --- if $R_H\geq 0$ then the QFEC certainly holds (but with an \emph{a priori} unknown value for the bound).

For the determinant we have
\begin{equation}
\det\langle T^{ab}\rangle_B = \rho_B \, p_B = {a_1^2\over4} \, {(b')^2\over1-b} \, \left[  b''  + {{1\over4}  (b')^2\over1-b}\right].
\end{equation}
This is positive iff
\begin{equation}
 b''  > - {{1\over4}  (b')^2\over1-b}.
\end{equation}
Certainly a sufficient condition for the determinant to be positive is $R>0$. 

For the trace-squared condition
\begin{equation}
\left(\langle T^{ab}\rangle\; \langle T_{ab}\rangle\right)_B = \rho_B^2 + \rho_B^2 > 0. 
\end{equation}
This is always satisfied, and provides no useful constraint.

For completeness we point out that near spatial infinity we can expand $b(x)$ in a Laurent series  $b(x) = b_1/x + b_2/x^2 + \dots$, and calculate
\begin{equation}
\rho_B = -{2a_1b_1 \over x^3} +\O(x^{-4});
\qquad
p_B = - {a_1 b_1^2\over4x^4}+\O(x^{-5}). 
\end{equation}

\subsection{Hartle--Hawking vacuum}
For the Hartle--Hawking vacuum:
\begin{eqnarray}
\rho_{HH} &=& -a_1 b'' + a_1 { \kappa^2-{1\over4}  (b')^2\over1-b};
\\
p_{HH}&=& a_1 {\kappa^2- {1\over4}  (b')^2 \over1-b}.
\end{eqnarray}
In~\cite{Prado:2013} we have announced that the FEC is generically violated for this vacuum. Let us study this fact in deeper detail. To analyse the FEC we must evaluate:
\begin{equation}
\rho_{HH}^2 - p_{HH}^2 = a_1^2\, b''  \left[ b''  -  2 \left( { \kappa^2-{1\over4}  (b')^2\over1-b} \right)\right]. 
\end{equation}
Consider the asymptotic regime for large $x$. If $b(x)>0$ in the asymptotic regime, then by the assumed asymptotic flatness eventually we must have $b'(x)<0$ and $b''(x)>0$. But then since asymptotically $b(x)\to0$, $b'(x)\to0$, and $b''(x)\to0$, in the asymptotic regime we have
\begin{equation}
\rho_{HH}^2 - p_{HH}^2 \sim - 2 a_1^2\, \kappa^2 \, b'' (x)  < 0.
\end{equation}
So if $b(x)>0$ the FEC is violated in the asymptotic regime. 
We can explicitly verify this as follows: Near spatial infinity expand $b(x)$ in a Laurent series  $b(x) = b_1/x + b_2/x^2 + \dots$, and calculate
\begin{equation}
\rho_{HH}^2-p_{HH}^2 = -{4a_1^2 b_1\kappa^2\over x^3}  + \O(x^{-2}).
\end{equation}
To the required level of accuracy we have
\begin{equation}
\rho_{HH}^2-p_{HH}^2 = -{2a_1^2 \,\kappa^2\, b''(x)}  + \O(x^{-2}),
\end{equation}
in agreement with the argument above. 
Conversely, suppose that asymptotically at large $x$ we have $b(x)<0$.  Then from the assumed asymptotic flatness $b''(x) < 0$ in the asymptotic region.  But at $x_H$ we have $b(x_H)=1$.  Therefore there will be some point $x_*$ where $b(x)$ takes on its minimum value. At that point
\begin{equation}
b'(x_*) = 0; \qquad b(x_*) < 0; \qquad b''(x_*) > 0.
\end{equation}
But now we have $ b''(x_*) > 0$ while at asymptotically large $x$  we have $b''(x) < 0$. Thus by the intermediate value theorem there will be some $x_{**} > x_*$ such that:
\begin{equation}
b''(x_{**}) = 0.
\end{equation}
Now at that point $x_{**}$ we have
\begin{equation}
(\rho_{HH}^2 - p_{HH}^2)|_{x_{**}} = 0.
\end{equation}
Furthermore in the vicinity of that point a Taylor series expansion yields
\begin{equation}
(\rho_{HH}^2 - p_{HH}^2)  =  -2 a_1 b'''(x_{**}) \; {\kappa^2-{1\over4}b'(x_{**})^2\over1-b(x_{**})} \; (x-x_{**}) + \O([x-x_{**}]^2) .
\end{equation}
 Generically the derivative will be nonzero and there will be FEC violations on one side or the other.

On the other hand, near-horizon we have
\begin{eqnarray}
\rho_{HH} &=& -{a_1 R_H\over2}  - {3 \over4} a_1 R'_H\; (x-x_H) + \O([x-x_H]^2); 
\\
p_{HH}&=& + {a_1 R_H\over2} + {1\over4} a_1 R'_H\; (x-x_H) + \O([x-x_H]^2);
\end{eqnarray}
and furthermore
\begin{equation}
\rho_{HH}^2-p_{HH}^2 = {1\over2} a_1^2 R_H R'_H \; (x-x_H) + \O([x-x_H]^2).\quad
\end{equation}
So the FEC is satisfied on the horizon, while the near-horizon validity/failure of the FEC is controlled by the sign of $(R_H R_H')$.
At intermediate distances, while the classical FEC generically fails, the weak form of the QFEC certainly holds (with an \emph{a priori} unknown value for the bound).

A quantum generalisation of the standard pointwise energy conditions is again of interest, since in this case the near horizon approximation leads to
\begin{equation}
\rho_{HH}=\O(1),\qquad\rho_{HH}+p_{HH} = \O(x-x_H).
\end{equation}
Therefore, the weak QWEC holds.

For the determinant we have
\begin{equation}
\det\langle T^{ab}\rangle_{HH} = \rho_{HH} \; p_{HH} = 
{a_1^2} \; {\kappa^2-{1\over4}(b')^2\over1-b} \; \left[  -b''  + {\kappa^2-{1\over4}  (b')^2\over1-b}\right].
\end{equation}
This quantity tends to $-{1\over4} a_1^2 R_H^2 < 0$ on the horizon, and to $a_1^2 \kappa^4 >0$ at spatial infinity. 
So the sign of the determinant definitely changes.
For the trace-of-square condition
\begin{equation}
\left(\langle T^{ab}\rangle\; \langle T_{ab}\rangle\right)_{HH} = \rho_{HH}^2 + \rho_{HH}^2 > 0. 
\end{equation}
This is always satisfied, and provides no useful constraint.

For completeness we point out that near spatial infinity we can expand $b(x)$ in a Laurent series  $b(x) = b_1/x + b_2/x^2 + \dots$, and calculate
\begin{equation}
\rho_{HH} = a_1\kappa^2 +{a_1\kappa^2 b_1 \over x} +\O(x^{-2});
\qquad
p_{HH} = a_1 \kappa^2 + {a_1 \kappa^2 b_1\over x}+\O(x^{-2}). 
\end{equation}

\subsection{Unruh vacuum}
For the Unruh vacuum we have
\begin{equation}
\rho_U = a_1\left(  -b'' + { {1\over2} \kappa^2 -{1\over4} (b')^2\over1-b}\right);
\end{equation}
\begin{equation}
p_U=  a_1\left(  { {1\over2} \kappa^2-{1\over4}  (b')^2 \over1-b}\right);
\qquad
f_U = {{1\over2} a_1 \kappa^2\over1-b}.
\end{equation}
Near the horizon we have
\begin{equation}
\rho_U = - {a_1\kappa\over4(x-x_H)} - {9\over16} a_1 R_H + \O(x-x_H);
\end{equation}
\begin{equation}
p_U=  - {a_1\kappa\over4(x-x_H)} + {7\over16} a_1 R_H + \O(x-x_H);
\end{equation}
\begin{equation}
f_U =  + {a_1\kappa\over4(x-x_H)} + {1\over16} a_1 R_H + \O(x-x_H);
\end{equation}
Thus the classical energy conditions are violated, in an unbounded way, as one approaches the horizon.
Now consider:
\begin{equation}
\{ (\rho+f)^2-(p +f)^2\}_U = \{ (\rho-p)(\rho+p+2f)\}_U  = \rho_{HH}^2-p_{HH}^2;
\end{equation}
\begin{equation}
\{ (\rho-f)^2-(p -f)^2\}_U = \{ (\rho-p)(\rho+p-2f)\}_U = \rho_B^2-p_B^2;
\end{equation}
and
\begin{equation}
\rho_U^2 - f_U^2 = \rho_{HH} \; \rho_B.
\end{equation}
Thus the FEC in the Unruh vacuum simultaneously exhibits, (and is largely determined by), properties of FEC in the Boulware and Hartle--Hawking states. 
Explicitly carrying out a  near-horizon calculation
\begin{eqnarray}
\{ (\rho+f)^2-(p +f)^2\}_U &=& {1\over2} a_1^2 \,R_H\, R' _H\,(x-x_H) + \O([x-x_H]^2); 
\\
\{ (\rho-f)^2-(p -f)^2\}_U &=& {a_1^2\,\kappa\, R_H\over x-x_H} + \O(1);
\\
\rho_{U}^2-f_{U}^2 &=& {a_1^2\,\kappa\, R_H\over 4(x-x_H)} + \O(1).
\end{eqnarray}
As long as $b'' _H \geq 0$, that is $R_H\geq 0$, these quantities will be bounded from below. So one has the same general situation: generic FEC violations but QFEC being satisfied under mild conditions. 

The determinant condition does not look particularly pleasant. Near horizon we easily calculate
\begin{equation}
\{\rho \; p - f^2\}_U = -{1\over8} a_1^2 (2 R_H^2 - \kappa \,R ' _H ) + \O(x-x_H). 
\end{equation}
This is sufficient to guarantee that the determinant is bounded at the horizon (and hence is bounded throughout the spacetime). Asymptotically one has
\begin{equation}
\{\rho \; p - f^2\}_U = -{a_1^2 \kappa^2 b_1 \over x^3} + \O(x^{-4}).
\end{equation}

The trace of square condition does not look particularly pleasant. Near horizon we easily calculate
\begin{equation}
\{\rho^2 +  p^2 - 2f^2\}_U = +{1\over4} a_1^2 (2 R_H^2 + \kappa\, R' _H ) + \O(x-x_H). 
\end{equation}
This is sufficient to guarantee that the trace of square is bounded at the horizon (and hence is bounded throughout the spacetime). Asymptotically one has
\begin{equation}
\{\rho^2 + p^2 - 2 f^2\}_U =-{ 2 a_1^2\kappa^2 b_1\over x^3}+ \O(x^{-4}).
\end{equation}
So again we see boundedness of these nonlinear energy conditions is generic.

For completeness we point out that near spatial infinity we can expand $b(x)$ in a Laurent series  $b(x) = b_1/x + b_2/x^2 + \dots$, and note that density, pressure, and flux all have the same asymptotic expansion:
\begin{equation}
\rho_U\sim p_U \sim f_U = {1\over2} a_1\kappa^2 +{a_1\kappa^2 b_1 \over 2x} +\O(x^{-2}).
\end{equation}
We summarize the overall situation in table \ref{T:1+1-generic}.

\begin{table}[htdp]
\caption{1+1 dimensional geometry: Generic situation}
\begin{center}
\begin{tabular}{||c||c|c|c|c||}
\hline
\hline
Name                           &  usual ECs & FEC   & det$(T)$ & tr$(T^2)$ \\
\hline
\hline
Hartle--Hawking            & bounded &   bounded & bounded &\cmark \\
\hline
Boulware                        & \xmark &  $R_H\geq 0$ &  bounded &\cmark \\
\hline
Unruh                            & \xmark &   $R_H\geq0$ &  bounded & bounded \\
\hline
\hline
\end{tabular}
\end{center}
\label{T:1+1-generic}
\end{table}%

\section{Discussion}\label{sD}

In this paper, we have developed several nonlinear point-wise energy conditions and their semiclassical generalisations, studying their fulfillment in different situations of particular interest. Those situations are: the Casimir vacuum, vacuum polarization in Schwarzschild spacetime (both in the general case and for a massless conformally coupled scalar field), and vacuum polarization in 1+1 QFT. In the last two cases, we have also paid attention to the distinct characteristics of the Hartle--Hawking, Boulware, and Unruh quantum vacuum states.

In the first place, we have considered the classical FEC, which has already proved useful in developing classical entropy bounds~\cite{Abreu:2011}, and its quantum generalisation, the QFEC~\cite{Prado:2013}.
After studying the sufficient constraints for the FEC/QFEC to be satisfied, depending on the type of stress energy tensor, we have been able to conclude that the FEC is satisfied for the Boulware vacuum state when a massless conformally coupled quantum scalar field is taken into account in Schwarzschild space. The QFEC in its weak formulation is fulfilled in all the examples we have considered. The only potential exceptions would be related, on one hand, with the Unruh vacuum state of a generic QFT in Schwarzschild space, since in this case we cannot extract any conclusion in general, and, on the other hand, with the 1+1 geometry for the Boulware and Unruh vacuum states if any interesting situation with negative curvature on the horizon could exist. Moreover, the strong formulation of the QFEC is satisfied in all the particular cases where we have been able to consider it.

In the second place, we have formulated and studied the DETEC and TOSEC. We have shown that those conditions are satisfied even in more situations than the FEC. In all considered cases we have been able to prove that the potential violations of those conditions are bounded, their quantum version therefore being satisfied. 
Furthermore, as the TOSEC is necessarily satisfied for stress energy tensors which are not of type IV, the only explicit example of its violation we have found is for the Unruh vacuum of a massless conformally coupled scalar field in Schwarzschild spacetime. 

It must be pointed out that although the DETEC and TOSEC have not as clear a physical interpretation as that of the FEC/QFEC, their formulation could still be of crucial interest. That is, the relevance of an energy condition would reside in the fact that it could be useful to extract consequences of spacetime where they are satisfied, if they are widely satisfied. Therefore, only a deep investigation of their consequences could tell us what is the most compelling approach.

On the other hand, we have also considered the possible formulation of the quantum generalisation of the standard point wise energy conditions. As we have shown, those generalisations would not be satisfied in general, with the Casimir vacuum and Hartle--Hawking vacuum being examples of particularly well-behaved situations from this point of view. Therefore, we can conclude that the key issue is the fact that the new energy conditions are nonlinear functions of the stress-energy. Nevertheless, it should be noted that whereas the nonlinearities are essential to keeping the relevant quantities bounded, they also imply that distinct QFTs could interfere in a potentially destructive manner. That is, the consideration of two QFT theories which the individual vacuum polarization effects satisfy the semiclassical nonlinear energy conditions, could at least in principle lead to a situation where those conditions are violated for the sum of the stress-energies. The behaviour of squeezed state excitations of the 
quantum vacuum might potentially be of significant interest in this regard~\cite{private}.

\section*{Acknowledgments}

PMM acknowledges financial support from the Spanish Ministry of Education through a FECYT grant, via the postdoctoral mobility contract EX2010-0854. 
MV was supported by the Marsden Fund, and by a James Cook fellowship, both administered by the Royal Society of New Zealand.  We wish to thank Chris Fewster, Larry Ford, and Tom Roman, for their comments and feedback.



\end{document}